\documentclass[aps,pre,10pt,twocolumn,superscriptaddress,nofootinbib]{revtex4-1}

\usepackage{amsmath,amssymb,amsfonts,mathtools,bm}
\usepackage{graphicx}
\usepackage{hyperref}
\usepackage{algorithm}
\usepackage{algorithmic}
\makeatletter
\renewcommand\ftype@algorithm{8}
\makeatother
\usepackage{xcolor}
\usepackage{booktabs}

\hypersetup{
colorlinks=true,
linkcolor=blue,
citecolor=magenta,
urlcolor=blue
}

\newcommand{\e}{\mathrm{e}}
\newcommand{\ii}{\mathrm{i}}
\newcommand{\Z}{\mathcal{Z}}
\newcommand{\F}{\mathbb{F}}
\newcommand{\wt}{\mathrm{wt}}

\begin{document}

\title{Computing Classical Partition Functions via Cycle-Space Sampling}

\author{Yoav Feigenbaum}
\affiliation{Department of Computer Science,
School of Information and Computer Sciences, University of California, Irvine,
Irvine, CA 92697, USA}

\author{Lev Barash}
\affiliation{Information Sciences Institute, University of Southern California,
Marina del Rey, California 90292, USA}

\author{Itay Hen}
\affiliation{Information Sciences Institute, University of Southern California, Marina del Rey, California 90292, USA}
\affiliation{Department of Physics and Astronomy, University of Southern California, Los Angeles, California 90089, USA}
\affiliation{Department of Electrical and Computer Engineering, University of Southern California, Los Angeles, California 90089, USA}


\begin{abstract}
\noindent We present a method for computing thermodynamic properties of classical
spin systems by reformulating the partition function as a sum over
even subgraphs of the interaction graph.  Starting from the
high-temperature expansion, we show that the expansion coefficients are
determined by the weight distribution of the null space of the binary
incidence matrix over~$\F_2$, and that this statement holds verbatim for
multi-spin interactions once the graph is replaced by a hypergraph.  
We introduce a Monte~Carlo scheme based on the $1/t$ Wang--Landau algorithm
that performs a random walk directly in this null space.  Because the
weight distribution is temperature independent, a single estimate yields
the zero-field partition function, and the thermodynamic quantities
obtained from its temperature derivatives, at arbitrary
temperature.  We validate the approach against exact finite-size results
on the two-dimensional triangular-lattice Ising model, apply it to the
three- and four-dimensional hypercubic Ising models where no exact
solution is available, and to the two- and three-dimensional Baxter--Wu
models, whose three-spin interactions place them outside the reach of
Pfaffian methods.
For the two-body models we compare against conventional
Wang--Landau sampling in energy space; for the Baxter--Wu models we
validate against exact enumeration at the smallest size in each
dimension.
\end{abstract}

\maketitle

\section{Introduction}
\label{sec:intro}

Partition functions play a central role in statistical mechanics,
encoding the complete thermodynamic content of a system.  For an
interacting spin model with Hamiltonian~$H$ defined on a set of
$N$~spins, the canonical partition function at inverse
temperature~$\beta = 1/(k_B T)$ is
\begin{equation}
\label{eq:Z-def}
\Z(\beta) = \sum_{\{s\}} \e^{-\beta H(s)},
\end{equation}
where the sum runs over all $2^N$ microscopic spin configurations.

Direct evaluation of~$\Z$ is generically intractable because the
configuration space grows exponentially with system size.  Analytical
solutions exist only for a handful of special cases, most notably
Onsager's solution of the two-dimensional Ising
model~\cite{Ising1925,Onsager1944,Kaufman1949}, its combinatorial
reformulation by Kac and Ward~\cite{KacWard1952}, and the
triangular-lattice result of Houtappel~\cite{Houtappel1950}; a broad
survey of the solvable cases is given by Baxter~\cite{BaxterBook}.  The
scarcity of such cases is not an accident of technique.  Computing the
Ising partition function is $\#$P-hard in
general~\cite{Valiant1979,JaegerVertiganWelsh1990}, and Barahona showed
that the associated decision problems are NP-hard already for the
three-dimensional lattice and for the two-dimensional lattice in a
field~\cite{Barahona1982}; Istrail later identified non-planarity as
the essential obstruction~\cite{Istrail2000}.  In the absence of exact
solutions, one
typically resorts to approximate methods:  Markov-chain Monte~Carlo
(MCMC) sampling~\cite{Metropolis1953,Swendsen1987}, series
expansions~\cite{Domb1974}, transfer-matrix
methods~\cite{Kramers1941}, or renormalization-group
approaches~\cite{Wilson1975}.

While MCMC methods can efficiently estimate thermal expectation values,
they face well-known challenges.  First, the estimation of the partition
function itself, as opposed to ratios of partition functions or thermal
averages, is notoriously difficult in the canonical
framework~\cite{Frenkel2002}.  Second, near critical points or in the
presence of frustration, standard local-update algorithms suffer from
critical slowing down, requiring increasingly long simulations to
decorrelate successive configurations~\cite{Sokal1997,LandauBinder}.
Extended-ensemble methods address the first difficulty by flattening
the sampled distribution---umbrella
sampling~\cite{TorrieValleau1977}, multicanonical
sampling~\cite{BergNeuhaus1992}, histogram
reweighting~\cite{FerrenbergSwendsen1988}, and replica
exchange~\cite{HukushimaNemoto1996}---and it is to this family that the
present work belongs.  Cluster
algorithms such as Swendsen--Wang~\cite{Swendsen1987} and
Wolff~\cite{Wolff1989} alleviate this problem for certain unfrustrated
models but do not extend straightforwardly to frustrated systems or
models with complex interactions, though specialized cluster moves
exist for some disordered cases~\cite{Houdayer2001} and for particular
multi-spin models such as the Baxter--Wu
model~\cite{NovotnyEvertz1993,VelonakisMartinos2011}.

A complementary approach is provided by high-temperature series
expansions, in which the partition function is expressed as a polynomial
whose coefficients count certain subgraphs of the interaction
graph~\cite{Domb1974,OitmaaHamerZheng,ButeraComi2002}.  The
high-temperature expansion of the Ising
model expresses~$\Z$ as a polynomial in $t \equiv \tanh(\beta J)$
weighted by the number of even subgraphs of each size; the
representation goes back to van der Waerden~\cite{vanDerWaerden1941}.
The central
challenge then shifts from sampling spin configurations to counting
combinatorial objects, a task amenable to a different set of
algorithmic tools.

In this work we develop this idea into a practical computational
framework.  The key observation is that the even subgraphs of a
graph~$G$ are in exact bijection with the elements of the null space of
the binary incidence matrix~$B$ of~$G$ over the field~$\F_2$.  The
high-temperature expansion coefficients are therefore determined by the
weight distribution of this null space, that is, the number of
null vectors of each Hamming weight.  We then introduce a Monte~Carlo
algorithm that performs a random walk directly in the null space,
employing the $1/t$ Wang--Landau scheme~\cite{Wang2001,WLt,Belardinelli2007}
to estimate the weight distribution.

Sampling in subgraph space rather than spin space is not a new idea,
and the distinctions bear stating precisely.  The
closest antecedent is the worm algorithm of Prokof'ev and
Svistunov~\cite{ProkofevSvistunov2001}, which samples the same
high-temperature even-subgraph representation by local moves that
introduce and displace a pair of odd-degree defects, at a temperature
fixed in advance.
Jerrum and Sinclair's fully polynomial randomized approximation scheme
for the ferromagnetic Ising partition function runs a Markov chain
whose states are subgraphs of the interaction graph rather than spin
configurations~\cite{JerrumSinclair1993}, and deterministic
polynomial-time approximations have since been developed for
bounded-degree graphs at parameters lying in a zero-free region of the
partition function~\cite{PatelRegts2017}.  Independently, the loop
calculus of Chertkov and Chernyak writes the exact partition function
as a belief-propagation term corrected by a series indexed by
generalized loops, subgraphs without dangling ends in which every
participating vertex has degree at least two, odd degrees
included~\cite{ChertkovChernyak2006a,ChertkovChernyak2006b}.  For the
zero-field pairwise Ising model expanded about the spin-symmetric
belief-propagation fixed point, the terms containing odd-degree
vertices vanish and the surviving generalized loops are precisely the
even subgraphs above.  In that case the loop series is indexed by the
same combinatorial data whose weight distribution we set out to
compute; we gather that data by Hamming weight rather than evaluate it
term by term.  What distinguishes the
present approach from all three is that the walk is performed in the
null space itself, using its $\F_2$-linear structure, and that what it
returns is the entire temperature-independent weight distribution
rather than a quantity tied to one temperature.

Our construction has three motivating features.  The weight
distribution is temperature independent, so once determined it yields
the partition function, and the thermodynamic quantities obtained from
its temperature derivatives, at any temperature without further
simulation; in this respect it is equivalent to a complete energy-space
density of states, from which it can be computed exactly
(Sec.~\ref{sec:results-bwexact}).  The random walk takes place in
a vector space over~$\F_2$: the moves are XORs with basis vectors, they
preserve the null-space constraint exactly, and they connect any two
elements in at most~$n$ steps, where $n$ is the dimension of the null
space.  Finally, nothing in the derivation refers to the topology or the
uniformity of the interaction structure.  As we show in
Sec.~\ref{sec:multispin}, the same identity holds when two-body
couplings are replaced by $k$-body couplings, with the incidence matrix
of a graph replaced by that of a $k$-uniform hypergraph.  This last
point carries the method beyond the reach of Pfaffian techniques, and
does so generically rather than through a construction tailored to a
particular model; it is the setting of a substantial part of our
numerical work.

The remainder of this paper is organized as follows.
Section~\ref{sec:HTE} reviews the high-temperature expansion,
establishes notation, and derives the multi-spin generalization.
Section~\ref{sec:methods} sets out the method: the null-space
formulation, the bijection with even subgraphs, and the connection to
the weight enumerator of the cycle code (Sec.~\ref{sec:nullspace}); the
sampling algorithm (Sec.~\ref{sec:algorithm}); the thermodynamic
observables and the numerical precision required to evaluate them
(Sec.~\ref{sec:observables}); complexity and scaling
(Sec.~\ref{sec:complexity}); and extensions, together with the
limitations of the approach (Sec.~\ref{sec:extensions}).
Section~\ref{sec:results} presents
numerical results for five lattice families.
Section~\ref{sec:discussion} concludes.

\section{High-temperature expansion}
\label{sec:HTE}

\subsection{The Ising model}

We consider the zero-field Ising model on a graph $G = (V,E)$ with
vertex set~$V$ ($|V|=N$) and edge set~$E$ ($|E|=M$):
\begin{equation}
\label{eq:H-ising}
H = -\sum_{\{i,j\}\in E} J_{ij}\, \sigma_i \sigma_j ,
\end{equation}
where $\sigma_i \in \{+1,-1\}$.  Throughout the main development we take
uniform ferromagnetic couplings, $J_{ij}=J>0$; the inhomogeneous case is
discussed in Sec.~\ref{sec:extensions}.

\subsection{Derivation of the expansion}

The partition function reads
\begin{equation}
\label{eq:Z-product}
\Z = \sum_{\{\sigma\}}
\prod_{\{i,j\}\in E}
\e^{\beta J \sigma_i\sigma_j}.
\end{equation}
Using the identity
\begin{equation}
\label{eq:cosh-identity}
\e^{\beta J \sigma_i\sigma_j}
=
\cosh(\beta J)\bigl[1 + \sigma_i\sigma_j \tanh(\beta J)\bigr],
\end{equation}
valid for $\sigma_i\sigma_j = \pm 1$, the partition function becomes
\begin{equation}
\label{eq:Z-expanded}
\Z = (\cosh \beta J)^{M}
\sum_{\{\sigma\}}
\prod_{\{i,j\}\in E}
\bigl[1 + t\, \sigma_i\sigma_j \bigr],
\end{equation}
where $t \equiv \tanh(\beta J)$.

Expanding the product over edges, each edge $e = \{i,j\}$ contributes
either a factor of~$1$ or a factor of~$t\,\sigma_i\sigma_j$.  A given
term in the expansion corresponds to a subset $S \subseteq E$:
\begin{equation}
\label{eq:Z-subsets}
\Z = (\cosh \beta J)^{M}
\sum_{S \subseteq E}
t^{|S|}
\sum_{\{\sigma\}}
\prod_{\{i,j\}\in S} \sigma_i \sigma_j .
\end{equation}

\subsection{Spin summation and even subgraphs}

To evaluate the inner sum over spins, note that for each vertex~$i$ the
spin~$\sigma_i$ appears exactly $\deg_S(i)$ times in the product, where
$\deg_S(i)$ is the degree of vertex~$i$ in the subgraph induced by~$S$.
The spin sum factorizes:
\begin{equation}
\sum_{\{\sigma\}}
\prod_{\{i,j\}\in S} \sigma_i \sigma_j
=
\prod_{i\in V}
\sum_{\sigma_i=\pm 1} \sigma_i^{\deg_S(i)}.
\end{equation}
Since
\begin{equation}
\sum_{\sigma=\pm1} \sigma^d =
\begin{cases}
2, & d \text{ even},\\
0, & d \text{ odd},
\end{cases}
\end{equation}
the spin sum is nonzero if and only if every vertex has even degree
in~$S$.  Subgraphs satisfying this condition are the even
subgraphs of~$G$, equivalently the elements of its cycle space.
When the condition is met the spin sum equals~$2^N$.

Denoting by $k_l$ the number of even subgraphs with exactly $l$ edges,
the partition function becomes
\begin{equation}
\label{eq:Z-HTE}
{
\Z(\beta) = 2^N (\cosh \beta J)^{M}
\sum_{l=0}^{M} k_l\, t^{l} ,
}
\end{equation}
where $t = \tanh(\beta J)$ and $k_0 = 1$ (the empty subgraph).  The
problem of computing~$\Z$ is thus reduced to determining the
coefficients $\{k_l\}_{l=0}^{M}$, a purely combinatorial and
temperature-independent task.

\subsection{Multi-spin interactions}
\label{sec:multispin}

Equation~\eqref{eq:Z-HTE} is usually presented as a statement about
graphs, but the derivation never uses the fact that an interaction
involves exactly two spins.  Consider
\begin{equation}
\label{eq:H-multispin}
H = -J \sum_{a \in \mathcal{A}} \prod_{i \in a} \sigma_i ,
\end{equation}
where $\mathcal{A}$ is a collection of $M \equiv |\mathcal{A}|$ subsets
of~$V$, each of arbitrary size.  Writing $\sigma_a \equiv \prod_{i\in a}
\sigma_i$ and noting that $\sigma_a = \pm 1$ and hence $\sigma_a^2 = 1$,
the identity~\eqref{eq:cosh-identity} applies unchanged,
\begin{equation}
\e^{\beta J \sigma_a} = \cosh(\beta J)\,[1 + t\,\sigma_a],
\end{equation}
and expanding the product over $\mathcal{A}$ gives
\begin{equation}
\Z = (\cosh\beta J)^M \sum_{S\subseteq\mathcal{A}} t^{|S|}
\prod_{i\in V}\sum_{\sigma_i = \pm 1}
\sigma_i^{\,\mu_S(i)} ,
\end{equation}
where $\mu_S(i) \equiv |\{a \in S : i \in a\}|$ is the number of
interaction terms in~$S$ containing site~$i$.  The spin sum is again
nonzero precisely when every $\mu_S(i)$ is even, in which case it
equals~$2^N$.  Equation~\eqref{eq:Z-HTE} therefore holds verbatim, with
$k_l$ now counting subsets of $l$ interaction terms in which every site
appears an even number of times.

It follows that the entire
apparatus developed below, including the null-space bijection and the sampling
algorithm, transfers to multi-spin models with no modification other
than the replacement of the graph incidence matrix by the incidence
matrix of the corresponding hypergraph.  The Baxter--Wu
model~\cite{BaxterWu1973}, defined by three-spin interactions on the
elementary triangles of the triangular lattice, is the case we treat
numerically in Sec.~\ref{sec:results-bw}.  Pfaffian methods do not
apply to it.  Transfer matrices do, as for any finite-range model, at a
cost exponential in the transverse size, and cluster algorithms have
been devised for the Baxter--Wu model specifically, either by freezing
one sublattice so that the remaining spins interact through effective
two-spin couplings~\cite{NovotnyEvertz1993} or by building clusters from
triangles~\cite{VelonakisMartinos2011}; these rely on the particular
structure of that model.

Equation~\eqref{eq:Z-HTE}, together with its multi-spin extension,
converts the evaluation of~$\Z$ into a counting problem that is exact,
finite, and independent of temperature: the zero-field partition
function, and with it the zero-field thermodynamics of the model, is
carried by the integer sequence $\{k_l\}$.  This is a
reformulation rather than a solution.  The objects being counted are
exponentially numerous, so direct enumeration is confined to the
smallest instances and the coefficients must instead be estimated.  How
to do so is the problem addressed in the next section.

\section{Method}
\label{sec:methods}

Our method rests on two ideas.  The first is structural: the
coefficients of the high-temperature expansion are exactly the weight
distribution of the null space of a binary incidence matrix, so
counting even subgraphs becomes a question about a linear code.  The
second is algorithmic: that null space can be explored directly by a
random walk whose moves are additions of basis vectors, and its weight
distribution estimated by flat-histogram sampling.  We take these in
turn, then show how the resulting coefficients yield the partition
function and its temperature derivatives at every temperature, and what numerical
precision that step demands.  The section closes with the cost of the
scheme, the way its accuracy degrades as systems grow, and the classes
of model it does and does not reach.

\subsection{Null-space formulation}
\label{sec:nullspace}

\subsubsection{The binary incidence matrix}

Let $B$ be the $N \times M$ incidence matrix over~$\F_2$ of the
interaction structure.  Each column corresponds to an interaction term
$a \in \mathcal{A}$ and has nonzero entries in the rows of the sites it
couples:
\begin{equation}
B_{va} =
\begin{cases}
1, & v \in a, \\
0, & \text{otherwise},
\end{cases}
\end{equation}
with all arithmetic modulo~2.  For two-body models every column has
exactly two nonzero entries and $B$ is the ordinary incidence matrix of
a graph; for $k$-body models every column has~$k$ nonzero entries and
$B$ is the incidence matrix of a $k$-uniform hypergraph.  Nothing in what
follows depends on which of the two cases applies.

\subsubsection{Bijection with null vectors}

We represent a subset $S \subseteq \mathcal{A}$ by its characteristic
vector $\mathbf{x} \in \{0,1\}^M$.  The number of terms of~$S$ incident
on site~$v$ is
\begin{equation}
\mu_S(v) = \sum_{a} B_{va}\, x_a = (B\mathbf{x})_v
\end{equation}
over the integers, so the requirement that every site be covered an even
number of times is
\begin{equation}
\label{eq:null-condition}
B \mathbf{x} = \mathbf{0} \pmod{2},
\end{equation}
i.e., $\mathbf{x} \in \ker B \subseteq \F_2^M$.  For a graph, $\ker B$
is the cycle space of~$G$~\cite{Diestel2017}.

The number of terms in~$S$ equals the Hamming weight
$\wt(\mathbf{x}) \equiv \sum_{a} x_a$, whence
\begin{equation}
\label{eq:kl-null}
{
k_l = \bigl|\{\mathbf{x} \in \ker B \;:\; \wt(\mathbf{x}) = l\}\bigr|.
}
\end{equation}

\subsubsection{Dimension of the null space}

By rank--nullity over~$\F_2$,
\begin{equation}
\label{eq:nullity}
n \equiv \dim(\ker B) = M - \mathrm{rank}_{\F_2}(B),
\end{equation}
and the null space contains exactly $2^n$ elements.  For a connected
graph $\mathrm{rank}_{\F_2}(B) = N-1$, so $n = M - N + 1$, the circuit
rank of~$G$; with $c$ connected components, $n = M - N + c$.

For hypergraphs the rank is not fixed by connectivity alone and must be
computed.  We record the instance that arises below.  For the Baxter--Wu
interaction structure on an open $L\times L$ triangular patch,
$N = L^2$ and $M = 2(L-1)^2$, and
\begin{equation}
\label{eq:bw-rank}
\mathrm{rank}_{\F_2}(B) = N - 2, \qquad n = M - (N-2) = (L-2)^2 ,
\end{equation}
for every $L \ge 2$, in agreement with direct computation for all
$L \le 20$.  The left null space of~$B$ is spanned by the two
sublattice-parity vectors: each elementary triangle contains one site
of each of the three sublattices of the triangular lattice, so the
indicator vector of any pair of sublattices annihilates every column,
and a propagation argument along the rows of the patch shows that
there are no further left null vectors.  The nullity $(L-2)^2$ equals
the number of interior sites, and the null space has a natural
geometric basis: the six triangles surrounding an interior site cover
that site six times and each of its neighbors twice, so their
characteristic vector lies in $\ker B$, and these $(L-2)^2$
``hexagon'' vectors are linearly independent.  On a torus the counting
is different---every site then belongs to six triangles, and the rank
deficiency depends on whether the three-sublattice structure is
compatible with the boundary identifications, which requires
$3 \mid L$.

\subsubsection{Parity of the weight distribution}
\label{sec:parity}

For multi-spin models of odd interaction order the weight distribution
obeys a constraint that, among two-body models, holds only for
bipartite graphs.

\textbf{Lemma.}  If every interaction term $a\in\mathcal{A}$ has odd
cardinality, then every element of $\ker B$ has even Hamming weight;
equivalently, $k_l = 0$ for all odd~$l$.

\textbf{Proof.}  Sum all $N$ rows of~$B$ over~$\F_2$.  The entry in
column~$a$ of the sum is $|a| \bmod 2 = 1$, so the sum is the all-ones
vector $\mathbf{1}\in\F_2^M$.  Hence $\mathbf{1}$ lies in the row space
of~$B$, which is the dual code of $\ker B$, and therefore
$\mathbf{1}\cdot\mathbf{x} = \wt(\mathbf{x}) = 0 \pmod 2$ for every
$\mathbf{x}\in\ker B$. $\square$

The hypothesis fails for two-body models, where the row sum vanishes
identically and no constraint follows in general.  Bipartite graphs are
the exception, since all their cycles are even and their cycle spaces
contain only even-weight vectors; the cycle code of the triangular
lattice, by contrast, does contain odd-weight words, the shortest being
the weight-three triangles.  For the Baxter--Wu model, by contrast, half of
the $M+1$ histogram bins are empty by construction.  We use this in two
ways: the sampler bins only even weights, halving the number of bins and
hence the work required for the histogram to flatten, and the vanishing
of the odd-$l$ coefficients serves as a check that the basis and the
walk have been implemented correctly.

A related observation concerns the top of the support.  The all-ones
vector lies in $\ker B$ precisely when every site is covered an even
number of times.  On a periodic Baxter--Wu lattice every site belongs
to six elementary triangles and the support therefore extends to
$l = M$ exactly.  On the open patches simulated below this fails:
boundary sites belong to one, two, or three triangles, the all-ones
vector is not a null vector, and the maximum Hamming weight in the
null space falls short of~$M$.

The shortfall can be computed exactly.  Writing a null vector as a
combination of the $(L-2)^2$ hexagon generators, a triangle carries a
one precisely when an odd number of its corners is an interior site
selected by that combination, so $l_{\max}$ is the maximum, over
subsets~$S$ of the interior sites, of the number of triangles with an
odd number of corners in~$S$.  We solve this optimization exactly
for every $3 \le L \le 20$, writing it as a maximization over the
$(L-2)^2$ binary generator coefficients with one parity constraint per
triangle and using the constraint-programming solver CP-SAT of
OR-Tools~\cite{ORTools}, which terminates only once the objective value is matched by
a proven upper bound.  The result is
\begin{equation}
\label{eq:bw-lmax}
l_{\max} = 2L^2 - 8L + 12 = M - (4L-10),
\end{equation}
for $L=3,4$ and $6 \le L \le 20$; $L=5$ is exceptional, with
$l_{\max}=24$ rather than~$22$, in agreement with complete enumeration
of its $2^9$ null vectors.  We have no proof that
Eq.~\eqref{eq:bw-lmax} holds for all~$L$.  Over the certified range the
shortfall $M - l_{\max} = 4L-10$ is a boundary term, so $l_{\max}/M$
approaches unity from below.  We compare
Eq.~\eqref{eq:bw-lmax} with the observed support edges in
Sec.~\ref{sec:results-bw}.

\subsubsection{Computing a basis for the null space}

A basis $\{b_1, \ldots, b_n\}$ for $\ker B$ can be computed by Gaussian
elimination in $O(NM\min(N,M))$ operations, and the sparsity of~$B$ (two
nonzero entries per column for graphs, $k$ for $k$-uniform hypergraphs)
reduces this substantially in practice; for very large instances the
sparse $\F_2$ methods of Wiedemann~\cite{Wiedemann1986} are
preferable.

For graphs one may instead avoid elimination entirely: given a spanning
tree~$T$, each non-tree edge $e \notin T$ defines a unique fundamental
cycle~$C_e$ consisting of~$e$ together with the tree path joining its
endpoints, and the characteristic vectors of these cycles form a basis
for~$\ker B$~\cite{Diestel2017}.  We use this construction for all
two-body models reported below and Gaussian elimination for the
Baxter--Wu models.

\subsubsection{Connection to coding theory}
\label{sec:coding}

The null space $\ker B$ is a binary linear code of length~$M$ and
dimension~$n$; for a graph it is the cycle code
of~$G$~\cite{MacWilliams1977}, and its dual (the row space of~$B$) is
the cut-space code.  Such graphical codes were introduced and studied
in the 1960s by Bredeson and
Hakimi~\cite{BredesonHakimi1967,HakimiBredeson1968} and revisited later
by Jungnickel and Vanstone~\cite{JungnickelVanstone1997}; for a
connected graph the cycle code has parameters $[M,\,M-N+1,\,g]$, where
the minimum distance~$g$ is the girth of~$G$~\cite{Bollobas1998}.  The
girth therefore fixes the smallest~$l$ with $k_l \neq 0$, which is
three for the triangular lattice and four for the square lattice, and
supplies a further exactly known coefficient against which an estimate
can be checked.  In this language the $k_l$ are the coefficients of
the weight enumerator polynomial,
\begin{equation}
\label{eq:weight-enum}
W(x,y) = \sum_{\mathbf{c}\in\ker B} x^{M-\wt(\mathbf{c})}\,
y^{\wt(\mathbf{c})}
= \sum_{l=0}^{M} k_l\, x^{M-l}\, y^{l},
\end{equation}
and the reduced partition function is a section of it,
$\tilde{\Z}(t) = W(1,t)$.

Three consequences follow.  The MacWilliams
identity~\cite{MacWilliams1977} relates the weight enumerator of the
cycle code to that of the cut-space code, which is a duality relation
for the partition function and a nontrivial constraint on any estimate
of the~$k_l$.  Coding-theoretic bounds of Plotkin and
linear-programming type constrain the support and shape of the weight
distribution.  And the literature on weight-enumerator computation for
structured codes is a source of specialized algorithms for graphs with
symmetry.

The identification also sets expectations about worst-case difficulty.
Computing the weight distribution of a general binary linear code is
intractable: the associated decoding and minimum-weight problems are
NP-hard~\cite{BMvT1978}, computing the minimum distance is
NP-hard~\cite{Vardy1997}, and the wider complexity landscape is
surveyed by Barg~\cite{Barg1998}.  Cycle codes of lattice graphs are
highly structured rather than generic, so these results do not preclude
efficient treatment of the cases of interest, but they do rule out any
general-purpose polynomial-time algorithm, consistently with the
hardness results for the Ising partition function noted in
Sec.~\ref{sec:intro}.

\subsection{Monte Carlo sampling of the null space}
\label{sec:algorithm}

\subsubsection{Random walk on the null space}

Let $\{b_1, \ldots, b_n\}$ be a basis for $\ker B$.  Every null vector
has the unique representation
\begin{equation}
\label{eq:null-rep}
\mathbf{x} = \bigoplus_{i=1}^{n} a_i\, b_i ,
\qquad a_i \in \{0,1\},
\end{equation}
where $\oplus$ is addition modulo~2.  The elementary move selects a
basis vector uniformly at random and applies
\begin{equation}
\label{eq:move}
\mathbf{x} \rightarrow \mathbf{x}' = \mathbf{x} \oplus b_i ,
\end{equation}
flipping the coefficient~$a_i$ and preserving the null-space constraint
exactly.  The chain is irreducible, since any null vector is reachable
from any other in at most~$n$ moves.  Because every move flips exactly
one coefficient~$a_i$ of the basis expansion, the bare walk is periodic
with period two; the acceptance step of Sec.~\ref{sec:WL} introduces
rejections that remove the periodicity (equivalently, a lazy step may
be added).  Under uniform selection the uniform distribution over the
$2^n$ null vectors is stationary.

\subsubsection{The $1/t$ Wang--Landau estimator}
\label{sec:WL}

Flat-histogram sampling of the Hamming weight is achieved with the
Wang--Landau algorithm~\cite{Wang2001,WangLandau2001b}.  We use the
$1/t$ variant~\cite{WLt,Belardinelli2007}, which eliminates the
saturation of the error that afflicts the original
protocol~\cite{ZhouBhatt2005}, together with a
discovery-and-restart rule adapted to the fact that the support of the
weight distribution is not known in advance.  The symbol~$t$ in the
established name of that variant refers to simulation time; to avoid a
collision with $t=\tanh\beta J$ we write the simulation clock
as~$\tau$ throughout.

We maintain $s(l) \equiv \ln \hat{k}(l)$, an estimate of the logarithm
of the weight distribution, a visited flag $v(l)$, and a global
discovery flag $g(l)$, all indexed by $l \in \{0,\ldots,M\}$.  A
proposed move $\mathbf{x}\to\mathbf{x}'$ from weight~$l$ to weight~$l'$
is accepted with probability
\begin{equation}
\label{eq:WL-accept}
p_{\mathrm{acc}} = \min\!\left(1,\; \e^{\,s(l) - s(l')}\right).
\end{equation}
Since the proposal is symmetric, for a frozen estimate~$s$ this rule
satisfies detailed balance with respect to the distribution
$\pi(\mathbf{x}) \propto \e^{-s(\wt(\mathbf{x}))}$, which becomes the
flat-histogram target $\pi \propto 1/k_l$ as $s(l) \to \ln k_l$.  The
adaptive process itself, in which $s$ changes after every move, is not
a reversible chain with a fixed target; its convergence rests on the
vanishing of the modification
factor~\cite{WLt,Belardinelli2007}.  The estimate at the resulting
weight is incremented by the modification factor~$F$, and the visited
flag is set.

The protocol rules are as follows; Algorithm~\ref{alg:WL} states them
in full.  The walk starts from the empty subgraph, $\mathbf{x} =
\mathbf{0}$, so that the weight $l = 0$ used for normalization is
discovered from the outset.  If the proposed move reaches a
weight~$l'$ that has never been seen ($g(l')$ false), the support of the
distribution has just been extended and every accumulated $s(l)$ is
biased relative to it; we therefore set $g(l')$ true, move the walker
to~$\mathbf{x}'$, and restart the estimate: $s$ and $v$ are
reinitialized, $F$ is reset to unity, the move counter is reset, and
the run returns to its first phase.  In the first phase, whenever every
discovered weight has been visited since the last reset, $v$ is cleared
and $F \to F/2$.  Writing $\tau$ for the number of trial moves divided
by the number of discovered weights, once $F \le 1/\tau$ the run enters
its second phase, in which $F$ is set to $1/\tau$ after every move and
the visited flags are no longer used.  A weight first encountered in
the second phase triggers the same restart.  The run terminates only
once it is in the second phase and $\tau \ge \tau_{\max}$, so the first
phase may run beyond the nominal budget (Sec.~\ref{sec:complexity}).
On termination $s$ is normalized by $s(0)$ and weights never
discovered are assigned $k_l = 0$.  Discovery is not completeness: the
condition that every discovered weight has been visited does not
certify that every accessible weight has been found, and the exactly
known support edges and the identity $\sum_l k_l = 2^n$ serve as
\emph{a posteriori} checks of that.
  In the second phase the
statistical error decreases as~$\tau^{-1/2}$ without
saturating~\cite{Belardinelli2007}.  Flat-histogram
methods nonetheless carry an intrinsic limitation independent of the
update schedule.  For walks that change the sampled variable by a
bounded amount per step, the round-trip time grows at least as the
square of the number of bins, and faster when the walk meets
bottlenecks~\cite{Dayal2004}.  Basis moves can change the weight by as
much as $\wt(b_i)$, so that bound does not carry over verbatim, but
bottlenecks in the weight coordinate slow the null-space walk in the
same way.  Diagnostics based on the spectral
gap of the transition matrix in the sampled
variable~\cite{BarashFadeevaShchur2017} and massively parallel
implementations~\cite{Vogel2013} apply directly to the null-space walk,
and we return to both in Sec.~\ref{sec:discussion}.

The estimate is determined only up to an additive constant in~$s$, which
is fixed by the exactly known value $k_0 = 1$.  Where additional
coefficients are known analytically, for instance the smallest nonzero
weight and its multiplicity on a regular lattice, they provide an
independent check; we use $\sum_l k_l = 2^n$ for this purpose in
Sec.~\ref{sec:results}.

\subsubsection{Efficient weight-change computation}

The cost per step is dominated by evaluating the Hamming-weight change.
From
\begin{equation}
\wt(\mathbf{x} \oplus b_i) = \wt(\mathbf{x})
+ \wt(b_i) - 2\, \wt(\mathbf{x} \wedge b_i),
\end{equation}
where $\wedge$ is the bitwise AND, we have $\Delta l = \wt(b_i) - 2\,
\wt(\mathbf{x} \wedge b_i)$.  Since $\wt(b_i)$ is precomputed, the cost
is that of the inner product $\wt(\mathbf{x} \wedge b_i)$, evaluated
in~$O(M/w)$ time with word-level operations ($w$ the machine word size,
typically~64).  For sparse basis vectors one may instead iterate over
the support of~$b_i$, at cost $O(\wt(b_i))$, which is the better choice
for fundamental-cycle bases on large lattices.

\subsubsection{Basis preprocessing}

Mixing depends on the structure of the basis.  If many basis vectors
have overlapping supports the induced weight changes are correlated and
the walk decorrelates slowly.  The fundamental-cycle basis is a natural
default, since fundamental cycles of a breadth-first spanning tree have
localized support and small weight.  Alternatively one may apply random
$\F_2$ row operations to a given basis to decorrelate supports, at the
cost of increasing $\wt(b_i)$ and hence the per-step cost.

\begin{algorithm}[t]
\caption{$1/t$ Wang--Landau sampling in the null space.  The clock
$\tau$ is the number of trial moves per discovered weight.}
\label{alg:WL}
\begin{algorithmic}[1]
\REQUIRE Null-space basis $\{b_1,\ldots,b_n\}$; move budget $\tau_{\max}$
\STATE $\mathbf{x} \leftarrow \mathbf{0}$;\quad $l \leftarrow 0$
       \COMMENT{empty subgraph}
\STATE $s(\cdot),v(\cdot),g(\cdot) \leftarrow 0$;\quad
       $g(0), v(0) \leftarrow \mathrm{true}$
\STATE $F \leftarrow 1$;\quad $m \leftarrow 0$;\quad $\tau \leftarrow 0$;\quad
       $\mathrm{phase} \leftarrow 1$
\WHILE{$\mathrm{phase} = 1$ \textbf{or} $\tau < \tau_{\max}$}
  \STATE Choose $i \in \{1,\ldots,n\}$ uniformly at random
  \STATE $\mathbf{x}' \leftarrow \mathbf{x} \oplus b_i$;\quad
         $l' \leftarrow \wt(\mathbf{x}')$
  \IF{$g(l')$ is false}
    \STATE $g(l') \leftarrow \mathrm{true}$;\quad
           $\mathbf{x} \leftarrow \mathbf{x}'$;\quad $l \leftarrow l'$
    \STATE $s(\cdot), v(\cdot) \leftarrow 0$;\quad $v(l)\leftarrow\mathrm{true}$;\quad
           $F \leftarrow 1$;\quad $m \leftarrow 0$;\quad
           $\mathrm{phase} \leftarrow 1$ \COMMENT{restart}
  \ELSE
    \IF{$\mathrm{rand}() < \min(1,\, \e^{\,s(l)-s(l')})$}
      \STATE $\mathbf{x} \leftarrow \mathbf{x}'$;\quad $l \leftarrow l'$
    \ENDIF
    \STATE $s(l) \leftarrow s(l) + F$;\quad $v(l)\leftarrow\mathrm{true}$
  \ENDIF
  \STATE $m \leftarrow m+1$;\quad
         $\tau \leftarrow m / \#\{l: g(l)\}$
         \COMMENT{moves per discovered weight}
  \IF{$\mathrm{phase} = 1$}
    \IF{$v(l)$ true for all discovered $l$}
      \STATE clear $v(\cdot)$;\quad $F \leftarrow F/2$
    \ENDIF
    \IF{$F \le 1/\tau$}
      \STATE $\mathrm{phase} \leftarrow 2$
    \ENDIF
  \ENDIF
  \IF{$\mathrm{phase} = 2$}
    \STATE $F \leftarrow 1/\tau$
  \ENDIF
\ENDWHILE
\STATE Normalize: $s(l) \leftarrow s(l) - s(0)$ for discovered $l$
\RETURN $k_l = \e^{s(l)}$ if $g(l)$, else $k_l = 0$, for $l = 0,\ldots,M$
\end{algorithmic}
\end{algorithm}

\subsection{Thermodynamic observables}
\label{sec:observables}

Once $\{k_l\}$ is known, the zero-field partition function and the
thermodynamic quantities obtained from its temperature derivatives
follow analytically as functions of temperature.

\subsubsection{Partition function and free energy}

Defining the reduced partition function
\begin{equation}
\label{eq:Ztilde}
\tilde{\Z}(t) \equiv \sum_{l=0}^{M} k_l\, t^l,
\end{equation}
so that $\Z = 2^N (\cosh\beta J)^M \tilde{\Z}(t)$, the Helmholtz free
energy is
\begin{equation}
\label{eq:F}
F = -k_B T \Bigl[ N \ln 2 + M \ln(\cosh\beta J)
    + \ln \tilde{\Z}(t) \Bigr].
\end{equation}

\subsubsection{Internal energy and specific heat}

From $U = -\partial_\beta \ln\Z$, using $\partial_\beta t = J(1-t^2)$
and $\partial_\beta \ln\cosh\beta J = Jt$,
\begin{equation}
\label{eq:U}
U = -MJt - J(1-t^2)\,\frac{\tilde{\Z}'(t)}{\tilde{\Z}(t)},
\end{equation}
with $\tilde{\Z}'(t) = \sum_l l\, k_l\, t^{l-1}$.  The specific heat,
$C_V/k_B = -\beta^2 \partial_\beta U$, is
\begin{align}
\label{eq:CV}
C_V/k_B = \beta^2 J^2 \Biggl[
&M(1-t^2) + (1-t^2)^2
\frac{\tilde{\Z}''(t)}{\tilde{\Z}(t)} \nonumber \\
&- (1-t^2)^2
\left(\frac{\tilde{\Z}'(t)}{\tilde{\Z}(t)}\right)^{\!2}
- 2t(1-t^2)\frac{\tilde{\Z}'(t)}{\tilde{\Z}(t)}
\Biggr].
\end{align}
The entropy follows from $S = (U-F)/T$.

\subsubsection{Numerical evaluation}
\label{sec:precision}

Equations~\eqref{eq:F}--\eqref{eq:CV} are innocuous as written but
demand care in practice.  The dynamic range of the $k_l$ is enormous:
for the three-dimensional lattices of Sec.~\ref{sec:results-hd} the
weight distribution spans some five hundred orders of magnitude, far
outside double precision.  All sums over the $k_l$ are therefore evaluated in
arbitrary-precision arithmetic and converted to log-space only at the
end.

Two features make the evaluation benign despite the dynamic range.  All
$k_l$ are nonnegative and, for ferromagnetic couplings, $t \in (0,1)$,
so every term in $\tilde\Z$, $\tilde\Z'$, and $\tilde\Z''$ is positive
and no cancellation occurs; the sums are as well conditioned as their
magnitudes allow.  And because the $k_l$ enter only through the ratios
$\tilde\Z'/\tilde\Z$ and $\tilde\Z''/\tilde\Z$, an error in the overall
normalization of $\{k_l\}$ cancels identically in~$U$ and~$C_V$.  What
propagates into the observables is not the normalization but the error
in the shape of the distribution, which is why we report relative
errors bin by bin rather than only for the total.  The specific heat
is more exposed to such errors than the energy.  Its second and third
terms in Eq.~\eqref{eq:CV} combine to
$(1-t^2)^2 t^{-2}\,[\mathrm{Var}_t(l) - \langle l\rangle_t]$, a
difference of moments of the thermal weight distribution
$p_t(l) = k_l t^l/\tilde\Z(t)$, and where $C_V$ is small compared with
the individual moments, as at low temperature, relative errors in the
shape of $\{k_l\}$ are amplified.  This statistical amplification
persists even when the sums are evaluated with sufficient numerical
precision.

The situation for antiferromagnetic couplings is addressed separately in Sec.~\ref{sec:limitations}.

\subsubsection{Magnetization and susceptibility}

In an external field the partition function acquires additional
structure.  The zero-field susceptibility follows from
$\chi = N\beta(\langle m^2\rangle - \langle m\rangle^2)$ with
$m = N^{-1}\sum_i\sigma_i$.  For the two-body models in zero field,
and for the two-dimensional Baxter--Wu patches, whose sublattice-pair
reversals [Eq.~\eqref{eq:bw-rank}] leave $H$ invariant and together
flip every site, $\langle m\rangle = 0$ on any finite lattice.  This
does not carry over to the three-dimensional Baxter--Wu system, whose
incidence matrix has full row rank (Appendix~\ref{sec:app-bw3d}) and
which therefore has no nontrivial spin-reversal symmetry.  There
products of interaction terms can reduce to a single spin; for the
three faces of a tetrahedron that share a vertex~$a$,
$(\sigma_a\sigma_b\sigma_c)(\sigma_a\sigma_b\sigma_d)(\sigma_a\sigma_c\sigma_d)
= \sigma_a$, so $\langle\sigma_a\rangle$ acquires a nonzero
contribution at every finite temperature.  In either case, incorporating
$\langle\sigma_i\rangle$ or $\langle m^2\rangle$ requires tracking
subsets in which one or two designated sites have odd incidence, that
is, a coset of $\ker B$ rather than the null space itself.  We defer
this to future work.

\subsection{Computational complexity and scaling}
\label{sec:complexity}

\subsubsection{Null-space dimension and graph structure}

The dimension $n$ controls both the size of the sampling space ($2^n$)
and, through~$M$, the number of histogram bins ($M+1$).  For a regular
lattice of coordination number~$z$ with two-body couplings and
periodic boundaries, $M = zN/2$
and
\begin{equation}
\label{eq:n-scaling}
n = \left(\frac{z}{2} - 1\right) N + 1 .
\end{equation}
For the open lattices simulated in Sec.~\ref{sec:results} the missing
boundary bonds reduce both counts,
\begin{equation}
M = \frac{z}{2}\,N - O\!\left(L^{d-1}\right), \qquad
n = \left(\frac{z}{2} - 1\right) N - O\!\left(L^{d-1}\right) + 1 ,
\end{equation}
leaving the leading behavior unchanged.  Table~\ref{tab:scaling}
collects the relevant cases; the ratios quoted there are
thermodynamic-limit ratios.  The entry that
matters for interpreting our results is the comparison in the last
column: $n/N$ is the asymptotic ratio of the dimension of the sampling space to
that of the space of spin configurations.  It equals unity for the
two-dimensional square lattice and for the two-dimensional Baxter--Wu
model, and exceeds it in every other case listed, reaching $n/N \to 7$
for the three-dimensional Baxter--Wu system.  A method that samples
the null space therefore operates in a space of higher dimension than
conventional energy-space Wang--Landau in most of the geometries
treated here, and a corresponding cost is to be expected.

\begin{table}[t]
\centering
\caption{Null-space dimension for the lattice families considered.
$z$ is the coordination number for two-body models and the number of
interaction terms per site for Baxter--Wu.  The last column is the ratio
of sampling-space dimension to spin-space dimension in the
thermodynamic limit.}
\label{tab:scaling}
\begin{ruledtabular}
\begin{tabular}{lccc}
Lattice & $z$ & $M$ & $n/N$ \\
\colrule
Square (2D)        & 4  & $2N$  & $1$ \\
Triangular (2D)    & 6  & $3N$  & $2$ \\
Simple cubic (3D)  & 6  & $3N$  & $2$ \\
Hypercubic (4D)    & 8  & $4N$  & $3$ \\
Baxter--Wu (2D)    & 6  & $2N$  & $1$ \\
Baxter--Wu (3D)    & 24 & $8N$  & $7$ \\
\end{tabular}
\end{ruledtabular}
\end{table}

\subsubsection{Cost per Monte Carlo step}
Each proposed null-space update costs $O(M/w)$ when implemented with
word-level operations, or $O(\wt(b_i))$ when the support of $b_i$ is
traversed sparsely.  By comparison, a single-spin update in the
energy-space implementation costs $O(d_i)$, where $d_i$ is the number
of interaction terms incident on spin $i$, and hence $O(1)$ for the
bounded-degree lattices considered here.  A null-space update may alter
$\wt(b_i)$ interaction variables simultaneously, so the two types of
updates are not directly comparable on a move-for-move basis.  Their
relative efficiency must instead be assessed using a common
computational budget together with autocorrelation or histogram
round-trip measurements.

The numerical comparisons reported in Sec.~\ref{sec:results} were
performed as follows.  The
null-space simulations used the nominal stopping parameter
$\tau_{\max}=10^9$ and were run on USC's high-performance computing
cluster, whereas the energy-space simulations used
$\tau_{\max}=10^7$ and were run on a workstation.  Moreover, in our
implementation the adaptive first stage of the $1/t$ Wang--Landau
procedure is allowed to continue until its discovery and histogram
conditions are satisfied before the $\tau_{\max}$ stopping condition
becomes operative.  Consequently, several energy-space runs performed
more than $10^7$ trial moves during this first stage, and the actual
number of moves varied with the model and system size.  The comparisons
below should therefore be interpreted as checks of consistency and
convergence under the computational budgets used, rather than as
controlled measurements of the relative computational efficiency of
the two methods.  A cost-normalized comparison at matched wall-clock
time remains necessary before any efficiency advantage can be claimed.

\subsubsection{Convergence}

Convergence depends on the number of histogram bins, $M+1$, and on the
dynamic range of $k_l$ across them.  For models of odd interaction order
the lemma of Sec.~\ref{sec:parity} halves the effective bin count at no
cost.  In the $1/t$ phase the statistical
error falls as $\tau^{-1/2}$ without the saturation of the original
protocol~\cite{Belardinelli2007}.  The practical difficulty at large
system size is not the asymptotic rate but the tails.  The extreme
weights $l \to 0$ and $l \to l_{\max}$, at the two ends of the
support, have very small $k_l$.  In a converged flat-histogram ensemble
they are visited as often as any other bin, but the walk reaches them
only through long excursions from the bulk, and the resulting
round-trip times leave their relative error orders of magnitude worse
than that at the mode.  This is visible in every relative-error panel
in Sec.~\ref{sec:results}.  How these errors reach the observables
depends on temperature through $p_t(l) = k_l t^l/\tilde\Z(t)$.  At high
temperature, $t \to 0$, the small-$l$ coefficients dominate; as the
temperature is lowered the weight moves toward the combinatorial bulk
of the distribution, since
$\partial\langle l\rangle_t/\partial\ln t = \mathrm{Var}_t(l) \ge 0$;
and the high-weight tail is suppressed by $t^l$ at every temperature.
At low temperature the errors in $C_V$ are therefore those of the bulk
coefficients, amplified as described in Sec.~\ref{sec:precision},
rather than those of the small-$l$ tail.

\subsubsection{Comparison with other methods}

Exact enumeration of all $2^n$ null vectors is feasible only for
$n \lesssim 30$.  Transfer-matrix methods compute $\Z$ exactly
for quasi-one-dimensional systems at cost exponential in the width.
Pfaffian methods of Fisher--Kasteleyn--Temperley
type~\cite{Fisher1961,Kasteleyn1961,Temperley1961} give polynomial-time
exact answers for planar graphs and do not extend beyond
them~\cite{Barahona1982,Istrail2000}.

Approximation schemes with guarantees exist in the ferromagnetic
case: the Jerrum--Sinclair chain is an FPRAS for any ferromagnetic
Ising system~\cite{JerrumSinclair1993}, and for bounded-degree graphs
deterministic approximations follow from zero-freeness arguments at
parameters inside a zero-free region~\cite{PatelRegts2017}.  These carry rigorous error bounds
that our scheme does not, but they estimate $\Z$ at a fixed temperature
rather than producing the temperature-independent~$\{k_l\}$, and they
are formulated for two-body couplings.
Multicanonical and simulated
tempering~\cite{Marinari1992} methods, and conventional Wang--Landau
sampling in energy space, operate in spin-configuration space and
estimate the density of states $g(E)$ rather than $k_l$.  For uniform
couplings the two sequences are related by an exact linear transform
[Eq.~\eqref{eq:dos-to-kl} and its inverse], so the distinction lies in
the space that is sampled, not in the information obtained.  Energy-space
Wang--Landau sampling is our baseline for the two-body models of
Sec.~\ref{sec:results}.

\subsection{Extensions and limitations}
\label{sec:extensions}

\subsubsection{Inhomogeneous couplings}

For edge-dependent couplings $J_a$, defining $t_a \equiv
\tanh(\beta J_a)$,
\begin{equation}
\label{eq:Z-inhomogeneous}
\Z = 2^N \prod_{a} \cosh(\beta J_a)
\sum_{\mathbf{x}\in\ker B} \prod_{a:\, x_a=1} t_a .
\end{equation}
Binning by Hamming weight no longer suffices, since null vectors of
equal weight carry different products.  One remedy is to introduce the
continuous variable $\mathcal{E}(\mathbf{x}) \equiv -\sum_a x_a \ln t_a$
and perform Wang--Landau sampling with respect to~$\mathcal{E}$, in
direct analogy with energy-space Wang--Landau.  This sacrifices the
temperature independence that motivates the method, since
$\mathcal{E}$ depends on~$\beta$ through the~$t_a$, and we have not
implemented it.

\subsubsection{Potts and clock models}

The $q$-state Potts model admits a high-temperature expansion in which
the even-subgraph condition is replaced by a divisibility
condition~\cite{Wu1982}.  This suggests replacing $\F_2$ with
$\mathbb{Z}/q\mathbb{Z}$, but a complete extension requires more than
the substitution: oriented edges carrying $\mathbb{Z}/q\mathbb{Z}$-valued
currents, a conservation constraint at every site given by the kernel
of the oriented incidence matrix, and weights given by the
character-expansion coefficients of the interaction, which for clock
models depend on the current and not only on whether it vanishes.  For
composite~$q$ the kernel is a module rather than a vector space.  We
have not developed this extension and regard it as prospective.

\subsubsection{Sign of the couplings}
\label{sec:limitations}

For two-body models the restriction to $J > 0$ is not a matter of
convenience.  For $J < 0$ one has $t = \tanh(\beta J) < 0$, and
whenever the cycle code contains odd-weight words,
$\tilde\Z(t) = \sum_l k_l t^l$ becomes an alternating sum over
coefficients that span hundreds of orders of magnitude.  The result is
catastrophic cancellation: the value
of~$\tilde\Z$ can be smaller than its largest term by a factor
comparable to $2^{-n}$, so extracting it requires the $k_l$ to a
relative precision that no sampling scheme can supply at large~$n$.
This is structurally the same obstruction that afflicts fermionic and
frustrated quantum Monte Carlo~\cite{TroyerWiese2005}, where the cost
of resolving a small average sign likewise grows exponentially.
Deciding whether such a problem can be removed by a change of basis is
itself computationally hard~\cite{MarvianLidarHen2019}, though useful
easing is sometimes possible~\cite{Hangleiter2020,Hen2019}.

On bipartite lattices the difficulty is removable, since the gauge
transformation $\sigma_i \to -\sigma_i$ on one sublattice maps the
antiferromagnet onto the ferromagnet; equivalently, the cycle code then
contains only even weights and no negative term survives.  For uniform
antiferromagnetic two-body couplings on non-bipartite lattices, which
is to say precisely the geometrically frustrated cases, it is not.

Models in which every interaction has odd order are free of the
difficulty for either sign of a uniform coupling.  Since $k_l = 0$ for
odd~$l$ (Sec.~\ref{sec:parity}), $\tilde\Z(-t) = \tilde\Z(t)$ and every
surviving term is nonnegative.  Physically, global spin reversal
changes the sign of every odd-order interaction and maps uniform $J<0$
onto uniform $J>0$; both Baxter--Wu models are therefore sign free for
either sign of~$J$.  More generally, for couplings of uniform magnitude
and signs $(-1)^{b_a}$, a spin-sign gauge
$\sigma_i \to (-1)^{y_i}\sigma_i$ removes every negative coupling
precisely when
\begin{equation}
\label{eq:gauge}
B^{T}\mathbf{y} = \mathbf{b} \pmod 2
\end{equation}
is solvable, that is, when $\mathbf{b}$ lies in the row space of~$B$,
or equivalently when $\mathbf{b}\cdot\mathbf{x} = 0$ for every
$\mathbf{x}\in\ker B$.  Uniform antiferromagnetic couplings correspond
to $\mathbf{b} = \mathbf{1}$, and the bipartite and odd-order cases
above are instances in which $\mathbf{1}$ lies in the row space.  We make no claim that the present method addresses frustration,
and all results below are ferromagnetic.  Whether a resummation of
$\tilde\Z$ that exploits the structure of the cycle code can control the
cancellation is an open question that we regard as worth pursuing.

\section{Numerical results}
\label{sec:results}

We report results for five lattice families, chosen to separate three
distinct questions: whether the sampler reproduces exactly known
answers, how it behaves where no exact answer exists, and whether it
functions at all for interactions beyond two-body.  For the two-body
models the baseline for comparison is conventional Wang--Landau sampling of the
density of states $g(E)$ in spin-configuration space, run with the same
$1/t$ protocol and implemented in the same code base.  We refer to the
two schemes as null-space and energy-space sampling.

\subsection{Setup}
\label{sec:results-setup}

Table~\ref{tab:systems} lists the systems, their sizes, and the
resulting null-space dimensions.  All simulations reported in this
work---null-space and energy-space alike---use open (free) boundary
conditions; couplings are uniform and ferromagnetic, $J = 1$.  Bases are
constructed from breadth-first fundamental cycles for the two-body
models and by Gaussian elimination for Baxter--Wu.  Each data point is
the mean of 20 independent runs with distinct seeds. The number of moves performed by the algorithm depends on
both $\tau_{\max}$ and the number of states discovered for that model and size; the relative standard error (RSE) plotted
in the lower panels of the figures is the standard error over these runs
divided by the mean, expressed as a percentage.  Where an exact result
is available we additionally plot the relative deviation of the mean
from it.

\begin{table}[t]
\centering
\caption{Systems studied.  All lattices carry open boundary
conditions.  $N$ is the number of sites, $M$ the number of interaction
terms, and $n = M - \mathrm{rank}_{\F_2} B$ the null-space dimension:
$n = M - N + 1$ for the connected two-body graphs, $n = (L-2)^2$
for the two-dimensional Baxter--Wu patches [Eq.~\eqref{eq:bw-rank}],
and $n = M - N$ for the three-dimensional Baxter--Wu systems, whose
incidence matrices have full row rank at every size simulated
(Appendix~\ref{sec:app-bw3d}).}
\label{tab:systems}
\begin{ruledtabular}
\begin{tabular}{llrrr}
System & $L$ & $N$ & $M$ & $n$ \\
\colrule
Triangular 2D & 4  & 16   & 33   & 18 \\
              & 8  & 64   & 161  & 98 \\
              & 12 & 144  & 385  & 242 \\
              & 16 & 256  & 705  & 450 \\
\colrule
Cubic 3D      & 4  & 64   & 144  & 81 \\
              & 6  & 216  & 540  & 325 \\
              & 8  & 512  & 1344 & 833 \\
              & 10 & 1000 & 2700 & 1701 \\
\colrule
Hypercubic 4D & 2  & 16   & 32   & 17 \\
              & 3  & 81   & 216  & 136 \\
              & 4  & 256  & 768  & 513 \\
              & 5  & 625  & 2000 & 1376 \\
\colrule
Baxter--Wu 2D & 5  & 25   & 32   & 9 \\
              & 10 & 100  & 162  & 64 \\
              & 15 & 225  & 392  & 169 \\
              & 20 & 400  & 722  & 324 \\
\colrule
Baxter--Wu 3D & 3  & 27   & 88   & 61 \\
              & 4  & 64   & 270  & 206 \\
              & 5  & 125  & 608  & 483 \\
              & 6  & 216  & 1150 & 934 \\
\end{tabular}
\end{ruledtabular}
\end{table}

The largest null spaces here, $n = 1701$ for the three-dimensional
lattice and $n = 1376$ in four dimensions, correspond to $2^{1701}$ and
$2^{1376}$ configurations respectively.  For the two-dimensional
Baxter--Wu patches $n = (L-2)^2$, the number of interior sites
[Eq.~\eqref{eq:bw-rank}].  The three-dimensional Baxter--Wu geometry is
defined in Sec.~\ref{sec:results-bw} and Appendix~\ref{sec:app-bw3d};
there $M = 2(L-1)^2(4L-1)$ and, because the site--face incidence matrix
has full row rank at these sizes, $n = M - N = 7L^3-18L^2+12L-2$.  The
entries of Table~\ref{tab:systems} for that system are obtained from
the lattice generator and confirmed by direct row reduction of~$B$,
independently of the sampled histograms.

\subsection{Validation against exact results}
\label{sec:results-tri}

\begin{figure*}[t]
\centering
\includegraphics[width=0.72\textwidth]{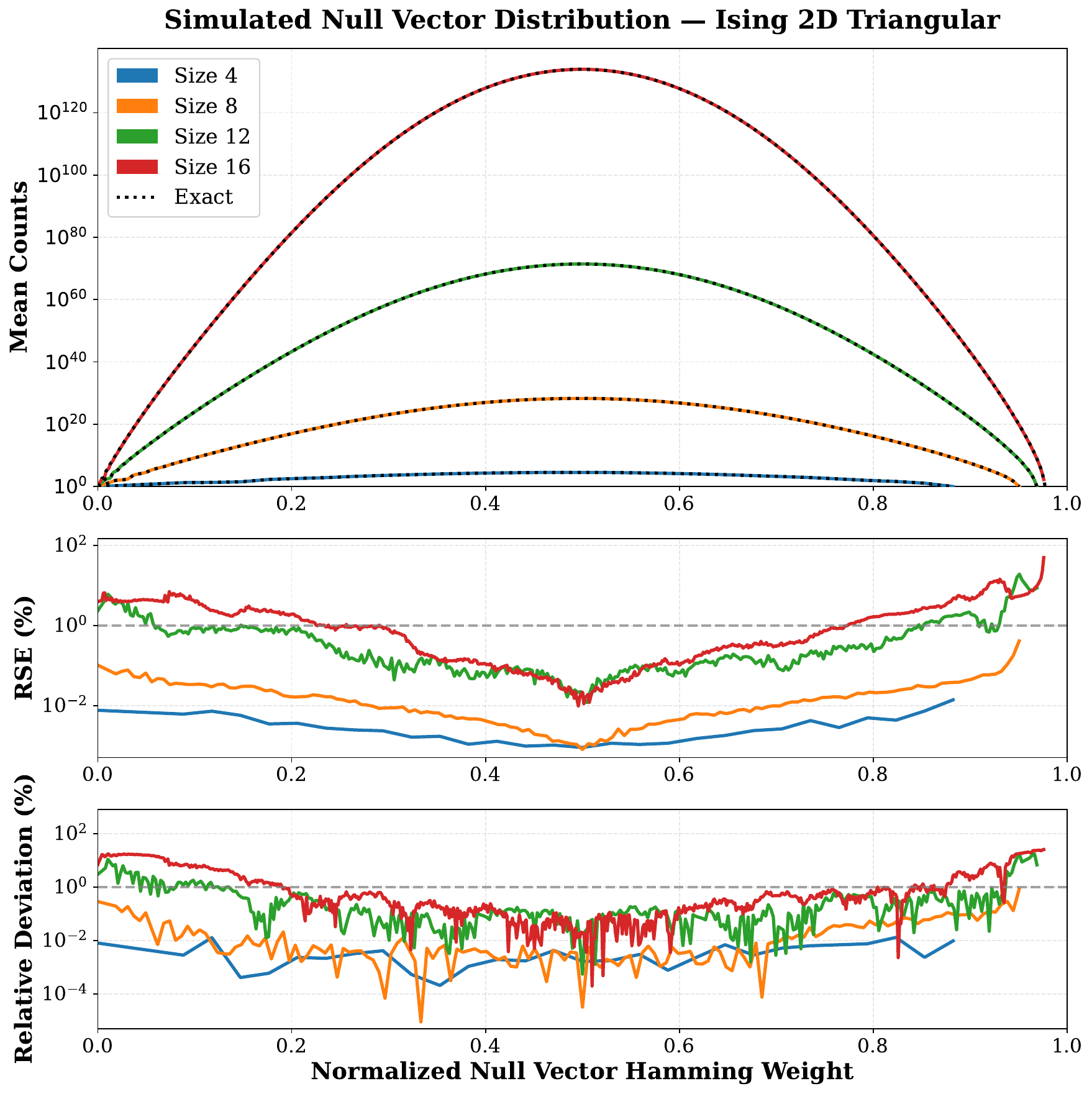}
\caption{Weight distribution of the cycle code of the $L\times L$
triangular lattice.  Top: mean estimated $k_l$ against normalized
Hamming weight $l/M$, with the exact enumeration shown as a dotted
black curve.  Middle: relative standard error over independent runs.
Bottom: relative deviation of the mean from the exact value.  The
horizontal axis of all three panels is $l/M$.  The error
is smallest at the mode $l/M \simeq 1/2$ and grows by several orders of
magnitude into both tails, where the coefficients are smallest and the
walk reaches them only through long excursions from the bulk.}
\label{fig:tri-null}
\end{figure*}

The two-dimensional triangular lattice is planar, so its finite-size
partition function is available exactly, and it provides an exact
reference at every simulated size.

The reference coefficients are obtained, at every~$L$, from a
Kac--Ward-type directed-walk determinant~\cite{KacWard1952,Cimasoni2010}
evaluated in exact arithmetic.  One builds the $6L^2 \times 6L^2$
transfer matrix~$M_1$ on directed edge states of the open patch, whose
nonzero entries are turning-angle phases and hence twelfth roots of
unity, and uses
\begin{equation}
\label{eq:det-square-main}
\det(I - v M_1) = G(v)^2, \qquad
G(v) \equiv \sum_{l=0}^{M} k_l\, v^l ,
\end{equation}
so that the generating polynomial of the $k_l$ is recovered as the
exact polynomial square root of a characteristic polynomial.  The
construction and its exact evaluation are given in
Appendix~\ref{sec:app-tri}.

Two features render this reference reliable for the present purpose.
First, $M_1$ is constructed on precisely the same finite graph---the
open $L\times L$ patch with $M = 3L^2-4L+1$ bonds and no periodic
identification---from which the sampled null-space bases are built, so
the comparison is not confounded by a boundary mismatch.  Second, the
resulting coefficients satisfy the two independent identities
$k_3 = 2(L-1)^2$, the number of elementary triangles of the patch, and
$\sum_l k_l = 2^{M-N+1}$, which is the order of $\ker_{\F_2} B$ for a
connected graph.  The second identity is sensitive to the boundary
condition---evaluated against the nullity of the corresponding periodic
lattice it would fail by many orders of magnitude---and therefore
guards against a boundary mismatch as well as an arithmetic error.
Neither identity determines the graph uniquely, since distinct graphs
can share these invariants, so they are consistency checks rather than
a certificate of the geometry.  For $L \le 4$ the
same coefficients were checked against complete enumeration of
$\ker_{\F_2} B$.

Figure~\ref{fig:tri-null} shows the estimated weight distribution.  The
agreement with the exact values is visually complete on the logarithmic
scale of the top panel, which spans some $135$ decades at $L=16$.  The
informative panels are the lower two.  At $L=4$ the relative
standard error is of order $10^{-2}\%$ or below across the whole
support.  At $L=8$ it rises from about $10^{-3}\%$ at the mode to
$\sim\!0.1\%$ at the lower edge of the support and a few tenths of a
percent at the upper edge.  At $L=12$ and $16$ it is about $10^{-2}\%$
at the mode, a few percent at the lower edge, and reaches tens of
percent near the upper edge.  The relative deviation from the exact
result tracks the standard error closely, as is expected of an
estimator dominated by variance rather than by a systematic bias that
would persist across seeds.  The degradation into the tails is the
effect discussed in Sec.~\ref{sec:complexity}.  Through $p_t(l)$ the
low-weight tail enters the observables most strongly at high
temperature, while the high-weight tail, where the errors are largest,
is suppressed by $t^l$ at every temperature.

\begin{figure*}[t]
\centering
\includegraphics[width=0.78\textwidth]{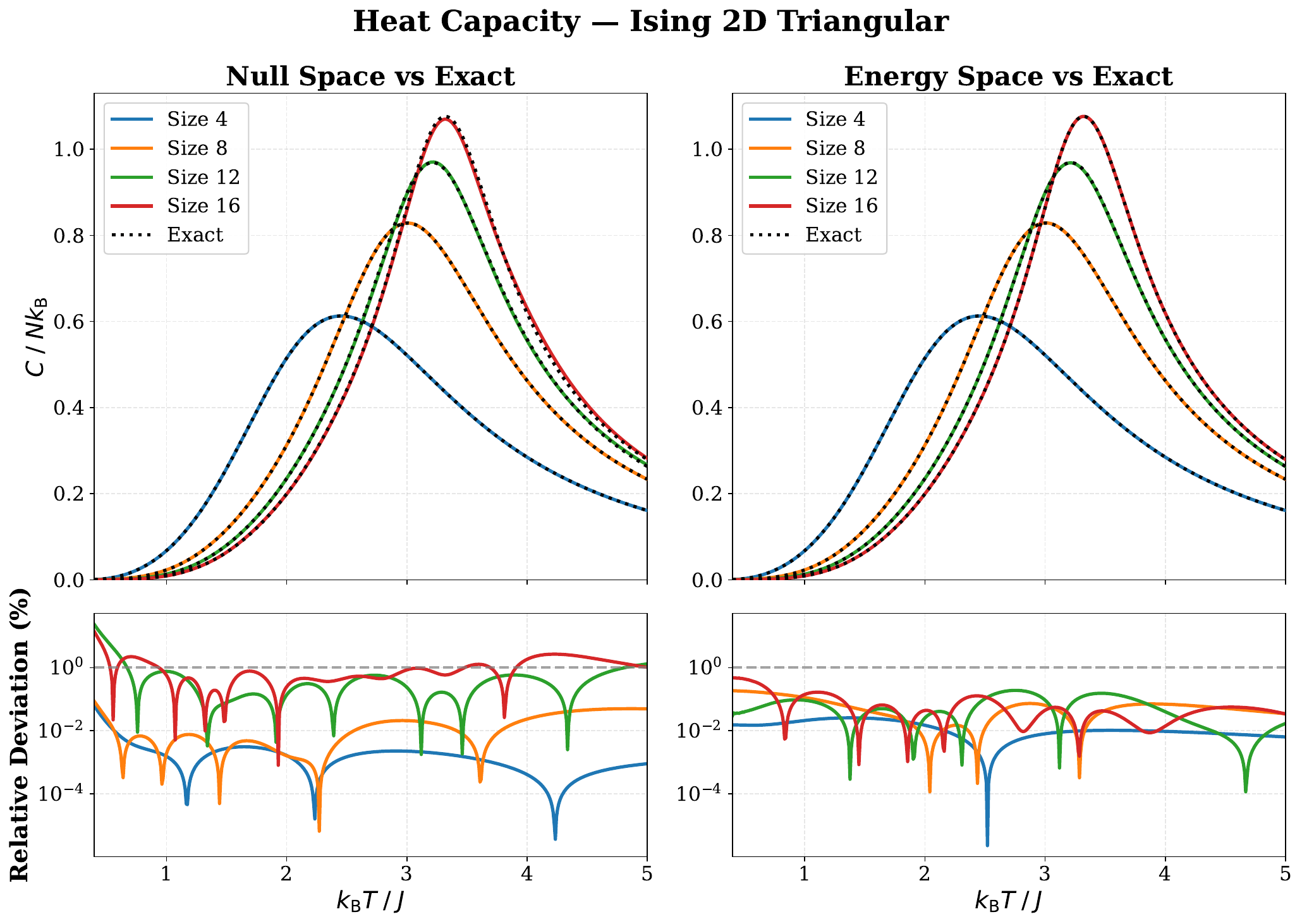}
\caption{Specific heat per site of the triangular-lattice Ising model
from cycle-space sampling (left) and from conventional Wang--Landau
sampling in energy space (right), each against the exact finite-size
result.  Lower panels show the relative deviation from exact.  The two
schemes agree with the exact answer at comparable accuracy, with the
energy-space estimator somewhat more uniform across temperature.}
\label{fig:tri-cv}
\end{figure*}

Figure~\ref{fig:tri-cv} propagates the weight distribution into the
specific heat through Eq.~\eqref{eq:CV} and compares against energy-space
Wang--Landau on the same lattices.  Both reproduce the exact finite-size
curves, including the growth and shift of the peak with~$L$.  The
deviation panels show cycle-space errors below $1\%$ over most of the
temperature range, exceeding it at $L=12$ and $16$ only near the ends
of the range shown, where they reach a few percent, with the
characteristic downward spikes where the deviation changes sign.  The
energy-space estimator is somewhat better behaved, holding below about
$0.5\%$ throughout.  We regard this as establishing correctness rather
than superiority: on a lattice where both methods apply and
$n \approx 2N$ (for the open patch $n = 2L^2-4L+2$), no accuracy
argument favors the null space, and we advance none.  The corresponding
mean-energy comparison is deferred to Fig.~\ref{fig:app-energy-tri} in
the Appendix, as is the energy-space density of states,
Fig.~\ref{fig:app-edist}.

\subsection{Three and four dimensions}
\label{sec:results-hd}

\begin{figure*}[t]
\centering
\includegraphics[width=0.48\textwidth]{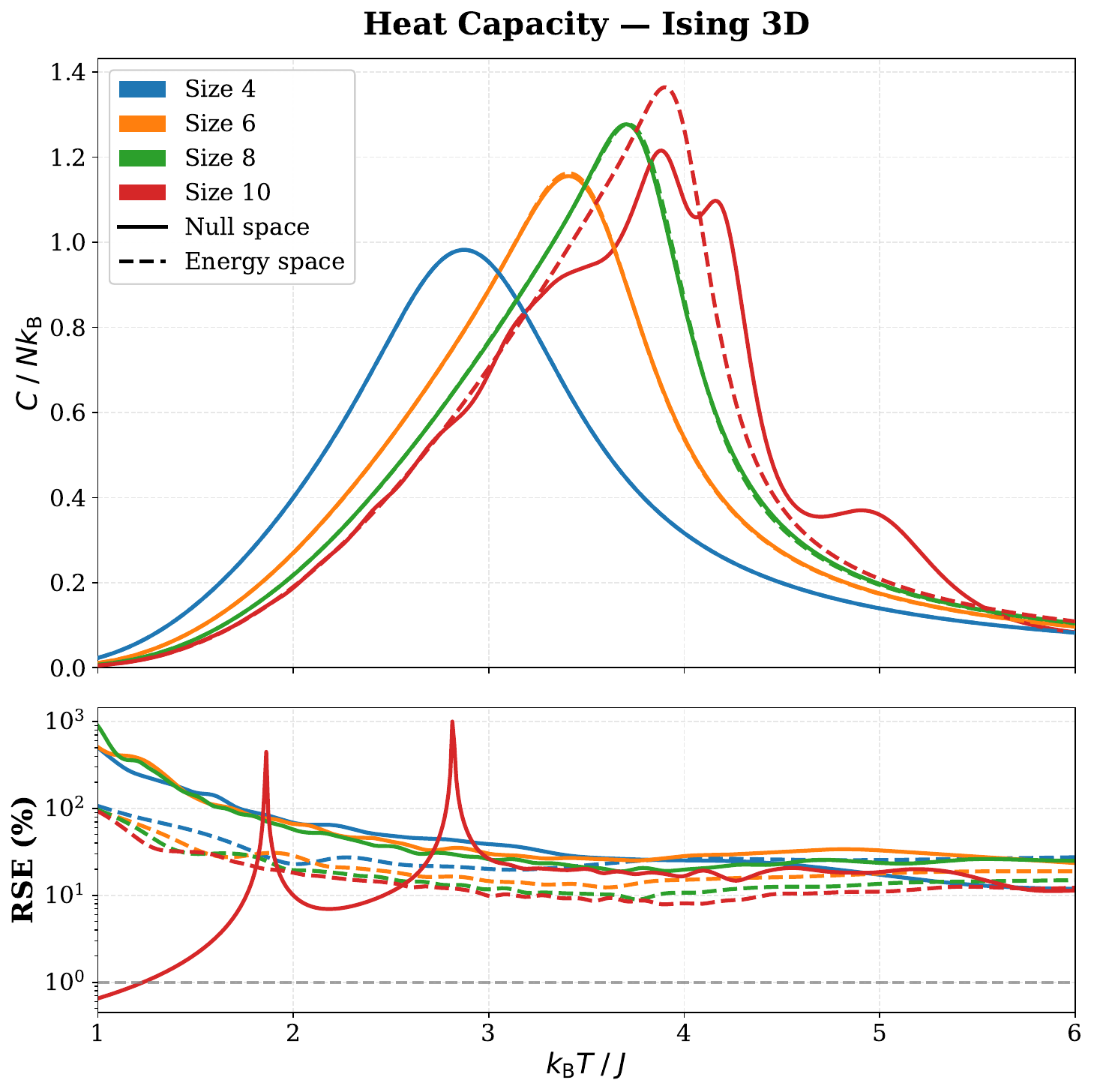}
\hfill
\includegraphics[width=0.48\textwidth]{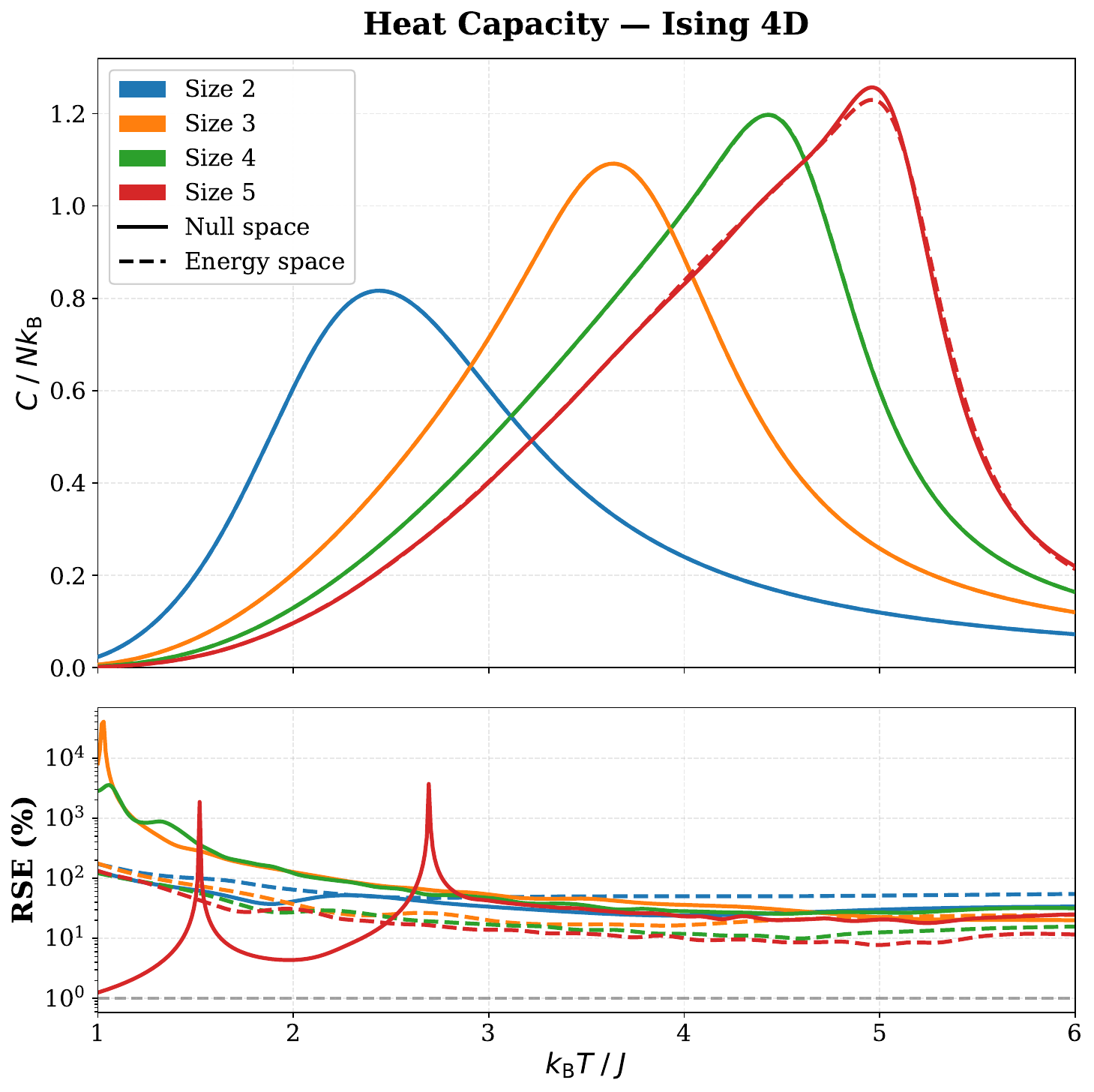}
\caption{Specific heat per site of the Ising model on the simple cubic
lattice (left) and the four-dimensional hypercubic lattice (right).
Solid curves are cycle-space estimates, dashed curves energy-space
estimates; lower panels give the relative standard error of each.
No exact finite-size reference was computed for either case.  The cycle-space $L=10$ curve
in the left panel has not converged: its shoulder near
$k_BT/J\simeq3.4$ and the two spurious maxima near $4.2$ and $4.9$,
absent from the energy-space result, are convergence artifacts and
carry no physical content.  It is shown to document the failure mode
and is excluded from the discussion of the physics.}
\label{fig:hd-cv}
\end{figure*}

Neither the three- nor the four-dimensional Ising model is exactly
solvable, and we did not compute exact finite-size references (which
would be feasible at the smallest sizes, e.g.\ the $2^{16}$
configurations of the four-dimensional $L=2$ system), so here the two
estimators check each other rather than an external reference.
Figure~\ref{fig:hd-cv} shows the specific heat.

At the smaller sizes the two schemes agree closely: the $L=4$, $6$, and
$8$ cubic-lattice curves lie on top of one another, and the peak
position and height evolve as expected toward the bulk critical
temperature; with open boundaries the finite-size corrections include
a surface term, so the approach to the bulk values is slower than for
periodic lattices.  The most precise Monte~Carlo determination gives
$K_c \equiv J/k_BT_c = 0.221\,654\,626(5)$, that is
$k_BT_c/J \approx 4.5115$~\cite{FerrenbergLandau1991,FerrenbergXuLandau2018}.
The four-dimensional curves behave similarly at $L=2,3,4$; there the
specific-heat singularity is marginal and its precise character remains
the subject of large-scale numerical
work~\cite{LundowMarkstrom2009,LundowMarkstrom2023}, so we make no
claim about the peak shape beyond its size dependence.

The relative standard errors in the lower panels of
Fig.~\ref{fig:hd-cv} are substantial for both estimators: roughly
$10$--$30\%$ above $k_BT/J \approx 3$, rising to $10^2\%$ and beyond
toward $k_BT/J = 1$, with the cycle-space values the larger.  The
agreement of the mean curves at $L=4$, $6$, and~$8$ described above
should be read with uncertainties of this size in mind.

At the largest size in each case the cycle-space estimate degrades
visibly, and in three dimensions it fails.  The $L=10$ cubic curve
develops a shoulder near $k_BT/J\simeq3.4$, a spurious secondary maximum
near~$4.2$, and a further bump near~$4.9$, none of which the
energy-space result reproduces; the corresponding relative standard
error panel shows sharp excursions to $10^2$--$10^3\%$ at
$k_BT/J\simeq1.9$ and $2.8$.  These are convergence artifacts, not
physical features: they appear only at the size where $n$ is largest,
and they are not stable across seeds.  We therefore do not regard the $L=10$
cycle-space curve as a result, and it should not be read as one; the
converged cycle-space data in three dimensions are those at
$L=4$, $6$, and~$8$.  We retain the curve in
Fig.~\ref{fig:hd-cv} only to make the failure mode visible, since
suppressing it would misrepresent the present state of the method.  At
nullities of order $10^3$ the $1/t$ protocol has not converged at the
budget used, and establishing convergence at these sizes requires either a longer budget or the
improvements sketched in Sec.~\ref{sec:discussion}.  In four
dimensions the $L=5$ mean curve agrees with the energy-space result,
but its error panel shows the same pattern of sharp excursions, near
$k_BT/J \simeq 1.5$ and $2.7$.

Across the range where both converge, the relative standard error of
the energy-space estimator is consistently below that of the
cycle-space estimator, typically by a factor of about two.  This is the expected consequence of
Table~\ref{tab:scaling}: at $z = 6$ and $z = 8$ the null space is
asymptotically two and three times the dimension of the spin space
($n/N = 1.70$ and $2.20$ at the sizes simulated), and the sampler
incurs the corresponding cost.  Temperature independence does not
distinguish the two methods here.  A complete $g(E)$ is equally
temperature independent, $\Z(\beta) = \sum_E g(E)\e^{-\beta E}$, and
for the uniform discrete models studied here its energy discretization
is exact, so neither estimate needs to be recomputed to change the
temperature.  What distinguishes cycle-space sampling is the
representation itself: the moves, the geometry of the space explored,
and the resulting numerical behavior.  In these geometries, at the
budgets used, that representation offers no accuracy advantage, and we
have not attempted a cost-normalized comparison.

The corresponding mean-energy and weight-distribution figures are
collected in the Appendix as Figs.~\ref{fig:app-energy-hd}
and~\ref{fig:app-null-hd}.

\subsection{Multi-spin interactions: the Baxter--Wu models}
\label{sec:results-bw}

\begin{figure*}[t]
\centering
\includegraphics[width=0.48\textwidth]{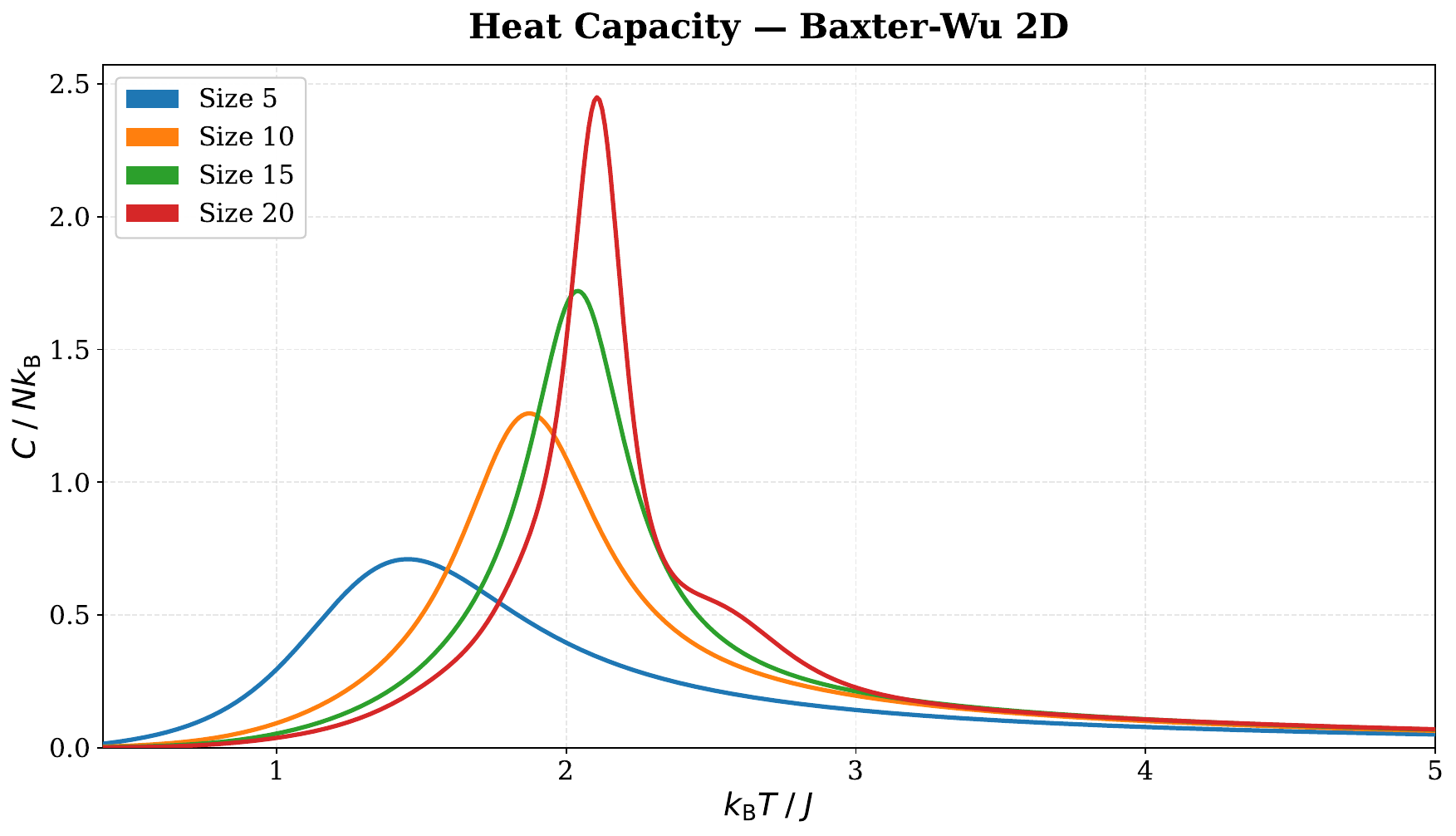}
\hfill
\includegraphics[width=0.48\textwidth]{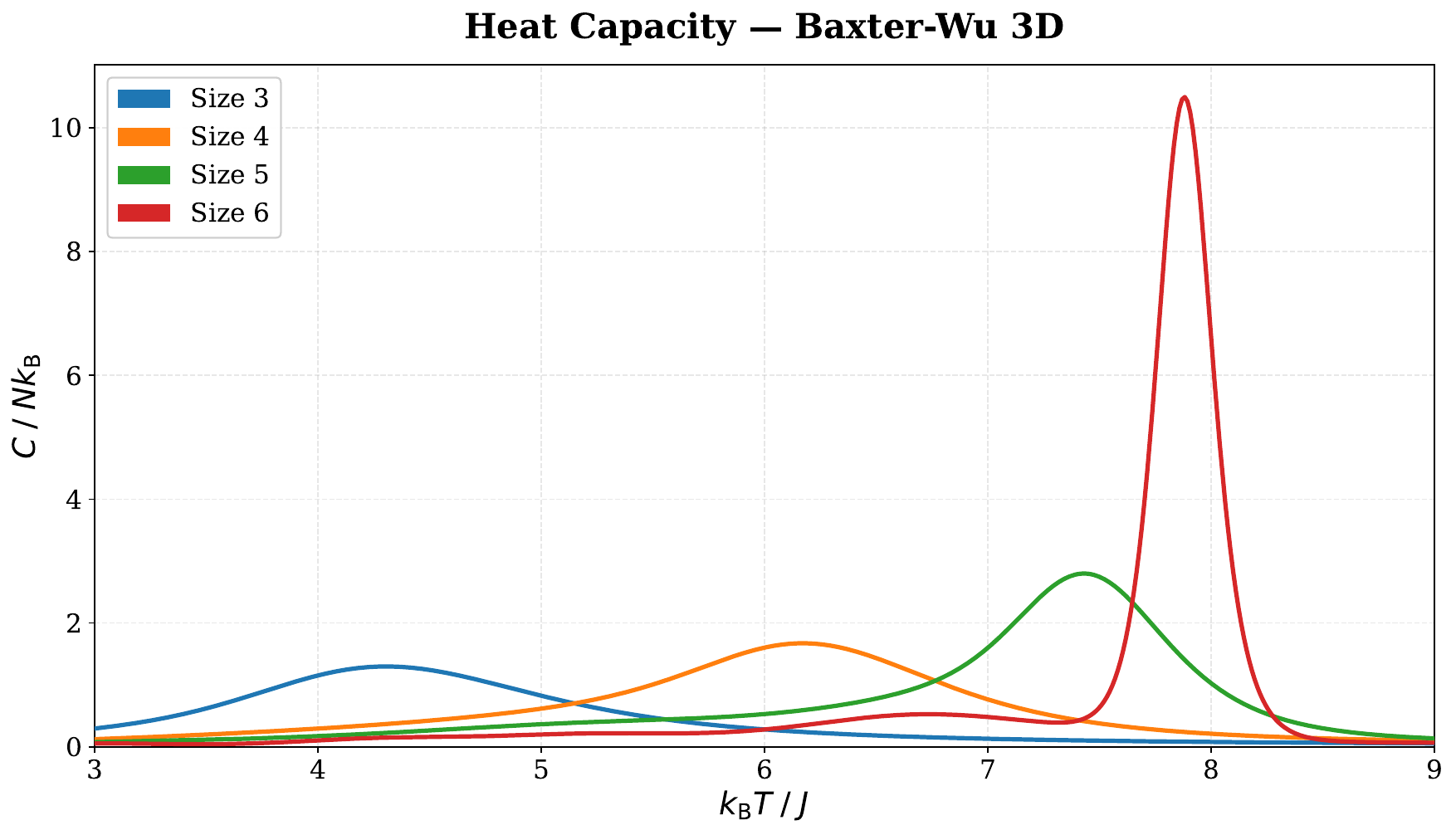}
\caption{Specific heat per site of the two-dimensional (left) and
three-dimensional (right) Baxter--Wu models from cycle-space sampling.
These models have three-spin interactions, so the sampling space is the
null space of a hypergraph incidence matrix.  No energy-space baseline
was run for these systems.}
\label{fig:bw-cv}
\end{figure*}

\begin{figure*}[t]
\centering
\includegraphics[width=0.48\textwidth]{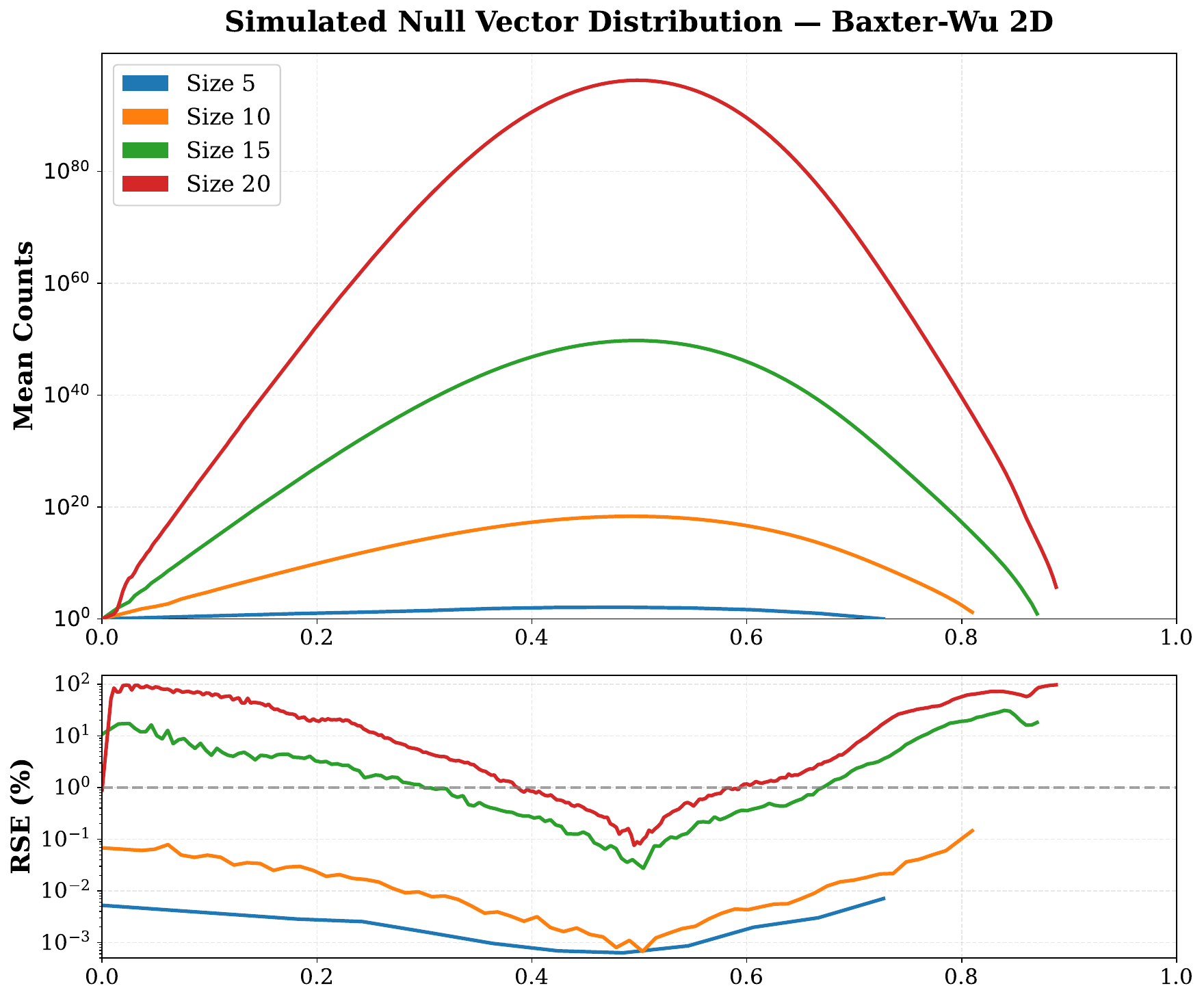}
\hfill
\includegraphics[width=0.48\textwidth]{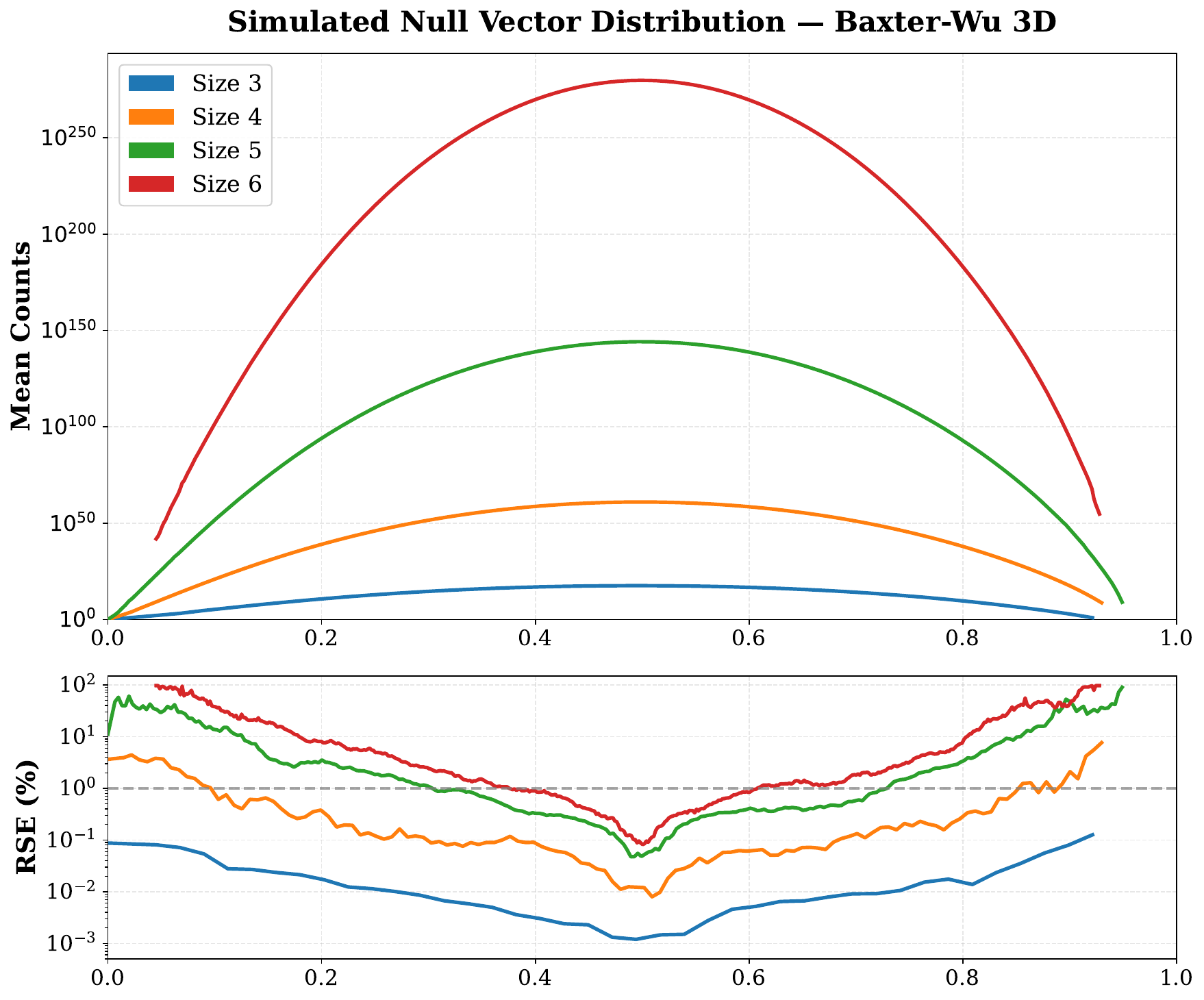}
\caption{Weight distributions for the two-dimensional (left) and
three-dimensional (right) Baxter--Wu models, with relative standard
errors below, plotted against the normalized Hamming weight $l/M$.
Only even weights are shown, since $k_l$ vanishes
identically for odd~$l$ in a model of odd interaction order
(Sec.~\ref{sec:parity}).  The curves terminate below $l/M = 1$; for
the open-boundary codes the support itself ends below~$M$ (exact
enumeration at $L=5$ gives $l_{\max}/M = 0.75$), as discussed in the
text.}
\label{fig:bw-null}
\end{figure*}

The Baxter--Wu model, introduced by Wood and
Griffiths~\cite{WoodGriffiths1972} and solved exactly by Baxter and
Wu~\cite{BaxterWu1973,BaxterWu1974}, places a three-spin coupling on
every elementary triangle of the triangular lattice.  It is self-dual,
with the same critical temperature as the square-lattice Ising model,
$k_BT_c/J = 2/\ln(1+\sqrt{2}) = 2.269\,185\ldots$, and belongs to the
universality class of the four-state Potts model, with
$\alpha = \nu = 2/3$ and $\gamma = 7/6$; unusually for that class, its
leading singularities carry no logarithmic
factors~\cite{BaxterWu1973,Macedo2023}.  Its ordered phase is fourfold
degenerate, reflecting the three-sublattice structure of the triangular
lattice, the same structure that underlies the rank result of
Eq.~\eqref{eq:bw-rank}.  The Pfaffian methods that solve planar
two-body models have no analogue here.  Cluster algorithms do exist,
but they are built on the particular structure of this model: Novotny
and Evertz freeze one sublattice, leaving effective two-spin couplings
on the other two~\cite{NovotnyEvertz1993}, and Velonakis and Martinos
construct clusters from triangles~\cite{VelonakisMartinos2011}.
Energy-space Wang--Landau sampling applies to multi-spin models as
well, and has been used for the three-dimensional
model~\cite{JorgeFerreiraCaparica2019}.  By Sec.~\ref{sec:multispin}
the cycle-space construction applies without modification and without
model-specific input.

\paragraph{The three-dimensional system.}
``Baxter--Wu 3D'' is not a standard model and admits more than one
natural definition, so we state ours explicitly.  We use an open finite patch of
the three-dimensional Baxter--Wu geometry of Jorge, Ferreira and
Caparica~\cite{JorgeFerreiraCaparica2019}, in which each bulk spin has
twelve nearest neighbors and belongs to twenty-four triangular faces,
and one three-spin interaction is placed on every elementary triangular
face of the nearest-neighbor graph.  Sites are labeled $(i,j,k)$ with
$0 \le i,j,k < L$, so $N = L^3$, and the twelve neighbor displacements
of an interior site are
\begin{equation}
\begin{split}
&(\bar 1,0,0),\ (0,\bar 1,0),\ (1,\bar 1,0),\
 (1,0,0),\ (0,1,0),\ (\bar 1,1,0),\\
&(0,1,1),\ (1,0,1),\ (0,0,1),\
 (\bar 1,0,\bar 1),\ (0,\bar 1,\bar 1),\ (0,0,\bar 1),\\ \\ 
\end{split} 
\label{eq:nn12}
\end{equation}
with $\bar 1 \equiv -1$.  A neighbor is retained only when all of its
coordinates remain in $\{0,\ldots,L-1\}$; no wraparound is applied.
Because the layer offsets above and below a given plane are negatives of
one another, the stacking is $ABC$, that is, face-centered cubic; the
lattice is described as close packed in
Ref.~\cite{JorgeFerreiraCaparica2019}, and a translation-invariant
twelve-vector neighbor set of this kind is necessarily the
face-centered variant.  The bulk geometry is thus the same as theirs,
while our finite systems differ by the use of free rather than periodic
boundaries.  Writing $\mathcal G_L = (V_L,E_L)$ for the resulting
nearest-neighbor graph, the interaction set is
$F = \{\{u,v,w\}\subset V_L : \{u,v\},\{v,w\},\{w,u\}\in E_L\}$, the
three-cycles of $\mathcal G_L$, which are exactly its elementary
triangular faces, and
$H_{\mathrm{BW3D}} = -J\sum_{\{u,v,w\}\in F}\sigma_u\sigma_v\sigma_w$.
Counting the faces by plane family (Appendix~\ref{sec:app-bw3d}) gives
\begin{equation}
\label{eq:bw3d-M}
M = 6L(L-1)^2 + 2(L-1)^3 = 2(L-1)^2(4L-1),
\end{equation}
and direct row reduction shows that the site--face incidence matrix has
full row rank for $L = 3,4,5,6$, so
$n = M - N = 7L^3 - 18L^2 + 12L - 2$, the values listed in
Table~\ref{tab:systems}.  The $E_{\mathrm{nn}} = 3L(L-1)(2L-1)$
two-body edges of $\mathcal G_L$ serve only to define the faces; they
are not themselves interaction terms.  The count that enters the
partition function, and hence the prefactor~$M$ in Eqs.~\eqref{eq:U}
and~\eqref{eq:CV}, is the number of faces, Eq.~\eqref{eq:bw3d-M}.  The
same caution applies in two dimensions, where $M = 2(L-1)^2$ is the
number of triangles and not the number $3L^2-4L+1$ of bonds of the
underlying triangular patch.  No exact result is available for the
three-dimensional model in the thermodynamic limit; for periodic
lattices with $L = 8,10,14,16,20$,
Ref.~\cite{JorgeFerreiraCaparica2019} reports a first-order transition
at $k_BT_c/J = 11.377\,485(29)$, with a specific-heat maximum scaling
as~$L^3$.

\paragraph{Specific heat.}
Figure~\ref{fig:bw-cv} shows the specific heat for the two-dimensional
model at $L = 5,10,15,20$ and for its three-dimensional counterpart at
$L = 3,4,5,6$.  The two-dimensional peaks sharpen with system size and
drift upward.  At $L=5$, where the weight distribution is known exactly
(Sec.~\ref{sec:results-bwexact}), the maximum is $C/Nk_B = 0.710$ at
$k_BT/J = 1.454$; at the larger sizes the maxima are $C/Nk_B = 1.26$,
$1.72$, and $2.45$, at $k_BT/J = 1.87$, $2.04$, and $2.11$, for
$L = 10$, $15$, and $20$ respectively.  The drift is toward the exact
$k_BT_c/J = 2.269\ldots$ from below, as expected of a finite-size
specific-heat maximum in a model with $\alpha>0$, although the approach
is slow---open boundaries contribute a surface correction to the
finite-size shift---and we have not attempted a finite-size scaling
extrapolation.
The three-dimensional model shows a markedly larger and narrower peak.
At $L=3$, again exactly known, the maximum is $C/Nk_B = 1.298$ at
$k_BT/J = 4.306$; it moves to $C/Nk_B = 1.67$, $2.80$, and $10.5$, at
$k_BT/J = 6.17$, $7.43$, and $7.88$, for $L = 4$, $5$, and $6$.  The
growth of the maximum is far steeper than in two dimensions, and the
$L=6$ peak is accompanied by a correspondingly steep rise in the mean
energy (Fig.~\ref{fig:app-energy-bw}), which would be consistent with
the first-order transition reported for the periodic
system~\cite{JorgeFerreiraCaparica2019}.  Figure~\ref{fig:bw-cv} shows
means over the 20 runs without uncertainty bands, and we have not yet
quantified the run-to-run scatter of these maxima.  This matters most
at $L=6$, whose nullity $n=934$ lies between those at which the
cubic-lattice runs converged ($n=833$) and failed ($n=1701$); the $L=6$
peak height should accordingly be regarded as provisional.
  The direction of the drift is
consistent with the periodic-lattice value $k_BT_c/J = 11.377\,485(29)$
of Ref.~\cite{JorgeFerreiraCaparica2019}, but the sizes accessible here
are far too small, and the boundaries too influential, to support an
extrapolation or to resolve the order of the transition.

Figure~\ref{fig:bw-null} shows the underlying weight distributions,
plotted over even weights only in accordance with the lemma of
Sec.~\ref{sec:parity}.  The curves terminate at $l/M \approx 0.75$,
$0.81$, $0.87$, and $0.90$ for $L = 5,10,15,20$ respectively, below
$l/M = 1$.  This is a property of the open-boundary code, not a
shortcoming of the sampler.  As discussed in Sec.~\ref{sec:parity},
boundary sites belong to fewer than six triangles, so the all-ones
vector is not a null vector and the support does not extend to
$l = M$.  Exact enumeration at $L=5$ ($M=32$, $n=9$) gives the support
$\{0\}\cup\{6,8,\ldots,24\}$: the minimum nonzero weight is
six---the hexagon of any interior site---and the maximum is
$l_{\max} = 24$ with $k_{24} = 1$.  At the larger sizes
Eq.~\eqref{eq:bw-lmax} gives $l_{\max} = 132$, $342$, and $652$ for
$L = 10,15,20$, that is $l_{\max}/M = 0.8148$, $0.8724$, and $0.9030$,
each certified optimal as described in Sec.~\ref{sec:parity}.  The
observed terminations coincide with these values at every size, so the
sampler reaches the true top of the support, and the shortfall relative
to $l/M = 1$ is the boundary effect of Eq.~\eqref{eq:bw-lmax} rather
than a failure to explore the tail.

The largest-weight coefficients remain the smallest and the least well
resolved, but their influence on the thermodynamics is mild: they
enter $\tilde\Z(t)$ multiplied by $t^l$ with $t<1$, so the observables
are controlled by the well-sampled bulk of the distribution.

No energy-space baseline was run for either Baxter--Wu system, so the
mutual comparison between representations that anchors
Secs.~\ref{sec:results-tri} and~\ref{sec:results-hd} is not available
here.  In its place we validate the smallest system in each dimension
directly against complete enumeration, in
Sec.~\ref{sec:results-bwexact}.  The corresponding mean-energy curves
appear as Fig.~\ref{fig:app-energy-bw} in the Appendix.

\subsection{Direct validation against exact enumeration}
\label{sec:results-bwexact}

Both Baxter--Wu systems admit an exact reference at their smallest
simulated size, which we use to test the sampler in the multi-spin
setting as the triangular lattice does in the two-body setting.

In two dimensions, $L=5$ has $n=9$, so the entire null space can be
enumerated: all $2^9 = 512$ elements of $\ker_{\F_2} B$ give
\begin{equation}
\label{eq:bw2d-exact}
\begin{aligned}
&k_0 = 1,\; k_6 = 9,\; k_8 = 16,\; k_{10} = 31,\; k_{12} = 73,\\
&k_{14} = 113,\; k_{16} = 122,\; k_{18} = 93,\; k_{20} = 43,\\
&k_{22} = 10,\; k_{24} = 1,
\end{aligned}
\end{equation}
all other coefficients vanishing.  The multiplicity $k_6 = 9 = (L-2)^2$
is the number of interior-site hexagons, consistent with the geometric
basis of Sec.~\ref{sec:nullspace}.

In three dimensions the null space at $L=3$ is far too large to
enumerate ($n = 61$), but the space of spin configurations is not: $N = 27$, so all
$2^{27}$ configurations can be swept directly to give the exact density
of states $g(E)$.  The weight distribution then follows from the
inverse of the high-temperature transform,
\begin{equation}
\label{eq:dos-to-kl}
\sum_{l} k_l\, t^l
= 2^{-N}\sum_{P} h(P)\,(1+t)^{P}(1-t)^{M-P},
\end{equation}
where $h(P)$ is the number of configurations with $P$ of the $M$
interaction terms satisfied, so that $E = -(2P-M)J$.  The transform is
invertible: setting $y = (1+t)/(1-t)$,
\begin{equation}
\label{eq:kl-to-dos}
\sum_P h(P)\,y^P = 2^{N-M}\sum_l k_l\,(y-1)^l(y+1)^{M-l},
\end{equation}
so for uniform couplings the density of states and the weight
distribution carry the same information.  Expanding the
right-hand side of Eq.~\eqref{eq:dos-to-kl} in exact integer arithmetic yields the complete
sequence $\{k_l\}$ for the $L=3$ system.  It reproduces every structural
prediction of Secs.~\ref{sec:nullspace} and~\ref{sec:parity}
independently of any simulation: all $k_l$ are nonnegative integers,
$k_l = 0$ for every odd~$l$ as required by the lemma of
Sec.~\ref{sec:parity}, and $\sum_l k_l = 2^{61}$ exactly, confirming
the nullity entered in Table~\ref{tab:systems}.  The support runs from
$l_{\min} = 4$ to $l_{\max} = 82$, with $k_4 = 130$ and
$k_6 = 1627$.

Propagating Eqs.~\eqref{eq:bw2d-exact} and~\eqref{eq:dos-to-kl} through
Eqs.~\eqref{eq:U} and~\eqref{eq:CV} gives exact finite-size
thermodynamics for both systems.  In two dimensions at $L=5$ the
specific heat peaks at $C/Nk_B = 0.7100$ at $k_BT/J = 1.454$; in three
dimensions at $L=3$ it peaks at $C/Nk_B = 1.2983$ at
$k_BT/J = 4.306$.  The sampled estimates reproduce both references to
better than $0.01\%$ in $C(T)$ and in $\langle E\rangle(T)$ across the
whole temperature range shown.

Two features of the exact reference are worth recording, since both are
readily mistaken in a multi-spin model.  The ground-state energy
is $E_0 = -MJ$ with $M$ the number of interaction terms, so
$\langle E\rangle/NJ \ge -M/N$ at every temperature: $-1.28$ for
two-dimensional $L=5$ and $-3.26$ for three-dimensional $L=3$.  And
$\partial\langle E\rangle/\partial T = \mathrm{Var}(H)/k_BT^2 \ge 0$, so
the mean energy is nondecreasing in temperature and the specific heat
is nonnegative.  Both are violated if the prefactor~$M$ in
Eqs.~\eqref{eq:U} and~\eqref{eq:CV} is taken to be the number of
two-body bonds of the underlying nearest-neighbor graph rather than the
number of three-spin terms, which is why we state the distinction
explicitly in Sec.~\ref{sec:results-bw}.  Both inequalities are
inexpensive to evaluate and worth imposing on any output of the
method.

\subsection{Consistency checks}
\label{sec:results-checks}

Independent of any reference solution, the total $\sum_l k_l$ must
equal $2^n$ with $n$ given by Table~\ref{tab:systems}.  Because the
total is dominated by the coefficients near the mode, this is a
one-parameter check on the overall normalization and shape of the
distribution.  Table~\ref{tab:sumcheck} collects the results: in every
case $\ln\sum_l \hat k_l$ agrees with $n\ln 2$
to the precision displayed.  At one decimal place in the logarithm this
constrains the total only to within about $5\%$, and it is blind to
errors in the shape that leave the total unchanged, so it is a coarse
normalization check rather than a precise accuracy certificate.  The
check is nonetheless unforgiving of any gross mismatch between the
interaction structure assumed in the analysis and the one actually
simulated---evaluated against the nullities of the corresponding
periodic lattices, for example, the same totals would appear to fall
short by factors of $2^{63}$ to $2^{500}$---so it doubles as a check on
the lattice geometry, though not a unique certificate of it, since
distinct structures can share the same nullity.

\begin{table}[t]
\centering
\caption{Consistency check of the estimated weight distributions against
the exact identity $\sum_l k_l=2^n$.  The estimated logarithmic totals
are quoted to the precision available in the run summaries.}
\label{tab:sumcheck}
\begin{ruledtabular}
\begin{tabular}{lcrrr}
System & $L$ & $n$ & $n\ln 2$ & $\ln\sum_l \widehat{k}_l$ \\
\colrule
Triangular 2D & 16 & 450  & 311.9162  & 311.9 \\
Cubic 3D      & 10 & 1701 & 1179.0434 & 1179.0 \\
Hypercubic 4D & 5  & 1376 & 953.7705  & 953.8 \\
Baxter--Wu 2D & 20 & 324  & 224.5797  & 224.6 \\
Baxter--Wu 3D & 6  & 934  & 647.3995  & 647.4 \\
\end{tabular}
\end{ruledtabular}
\end{table}
\section{Summary and conclusions}
\label{sec:discussion}

We have presented a framework that reformulates the computation of the
Ising partition function as a sampling problem in the null space of the
binary incidence matrix, and shown that the reformulation survives the
passage from two-body to multi-spin interactions unchanged.  Like a
complete energy-space density of states, the weight distribution
$\{k_l\}$ is temperature independent, so a single estimate yields the
zero-field partition function, and the thermodynamic quantities
obtained from its temperature derivatives, at any temperature.  For
the uniform models treated here the two coefficient sequences carry
equivalent information, each being an exact linear transform of the
other [Eqs.~\eqref{eq:dos-to-kl} and~\eqref{eq:kl-to-dos}]; what
differs is the sampling problem.  The cycle-space walk moves through a
linear code by additions of basis vectors rather than through spin
configurations, it applies without structural modification to
arbitrary multi-spin interaction hypergraphs, and its output is
directly the weight distribution of that code, which makes
coding-theoretic tools available and admits extensions to multivariate
weight enumerators for models with several interaction classes.  Its
numerical conditioning and convergence behavior differ from those of
energy-space sampling, as the results above show.  Sparse
interaction structures whose nullity is well below the spin-space
dimension constitute a further promising setting that we have not yet
explored.

The numerical results delineate the regimes in which each
representation is effective.  Whenever conventional energy-space
Wang--Landau sampling is also applicable and both calculations have
converged, cycle-space and energy-space sampling produce mutually
consistent thermodynamic observables with broadly comparable accuracy.
On the triangular lattice the energy-space estimates are somewhat more
uniform, while in three and four dimensions the larger null-space
dimension leads to visible convergence loss at the largest sizes
studied.  For the multi-spin models, where no energy-space baseline was
run, the estimator was instead checked directly against complete
enumeration at the smallest size in each dimension.  Because the
simulations were not performed at matched computational cost, these
results do not establish an efficiency ordering between the two
methods.

Several directions follow naturally from the present results.  The most
immediate concerns are convergence diagnostics, parallelization, and
cost-normalized benchmarking.  The spectral-gap diagnostic of Barash
et al.~\cite{BarashFadeevaShchur2017} could replace the fixed move
budget with a principled convergence criterion for the null-space
walk.  In parallel, the Wang--Landau framework of Vogel
et al.~\cite{Vogel2013}, which distributes the sampled weight range
across multiple replicas, could directly address the slow exploration
of the tails observed for our largest lattices.  A controlled comparison
with energy-space sampling should additionally match wall-clock time or
another common computational budget, record the actual number of trial
moves and histogram round trips, and include the one-time cost of
constructing the null-space basis.  Such measurements are needed to
determine the crossover between the two representations.

The tail problem may also be reduced by changing the sampling objective.
Rather than estimating the entire weight distribution uniformly, one
could sample null vectors with probability proportional to $t^l$ and
combine the resulting ensemble with an appropriate normalization
estimator for $\widetilde{\Z}(t)$.  This would concentrate computational
effort near the weights that dominate at the temperature of interest,
although it would sacrifice the temperature-independent character of
the full $\{k_l\}$ calculation.  Complementary information is available
through the MacWilliams identity, which imposes an exact relation
between the cycle- and cut-space weight enumerators.  Incorporating this
constraint, or sampling the dual enumerator when it is more favorable,
may improve the reconstruction of poorly resolved tails.

The coding-theoretic formulation also points toward deterministic and
hybrid alternatives to flat-histogram sampling.  The weight distribution
can be obtained as a marginal of a graphical model on the Tanner graph
of $\ker B$, suggesting generalized belief
propagation~\cite{Yedidia2005} and tensor-network
contraction~\cite{LevinNave2007,NishinoOkunishi1996} for interaction
graphs with suitable structure.  This connection is particularly
natural because, for the zero-field pairwise model at the
spin-symmetric belief-propagation fixed point, the surviving terms of
the loop series are the same even subgraphs counted by the
coefficients $\{k_l\}$%
~\cite{ChertkovChernyak2006a,ChertkovChernyak2006b}.  A hybrid scheme
could therefore use belief propagation to capture the leading
contribution and cycle-space sampling to estimate the weight-resolved
corrections.

The principal conceptual obstacle remains the cancellation problem for
frustrated systems, those whose coupling signs cannot be removed by the
gauge condition of Eq.~\eqref{eq:gauge}, described in
Sec.~\ref{sec:limitations}.  Overcoming this obstruction, whether
through a change of representation, a partial resummation, or a
structure-dependent decomposition of the cycle code, is the most
important open question raised by the present work.  A more speculative
direction is to seek a corresponding construction for quantum spin
models, where the partition function involves a trace over
noncommuting operators.  A Suzuki--Trotter expansion or a
permutation-matrix representation may provide an enlarged configuration
space in which an analogous null-space structure can be identified.

By shifting the computational effort from spin configurations to the
combinatorics of the interaction hypergraph, and by identifying the
high-temperature coefficients with the weight enumerator of a linear
code, the present formulation makes tools from coding theory available
for computing classical partition functions.  The results establish
the soundness of this reformulation and demonstrate its applicability
to interaction structures beyond the reach of graph-specific methods.
Improved convergence control, dual-space algorithms, and
cost-normalized benchmarks will determine the regimes in which this
structural flexibility also translates into a computational advantage.

\begin{acknowledgments}
The authors acknowledge the Center for Advanced Research Computing (CARC) at the University of Southern California for providing the computing resources used in this work.
\end{acknowledgments}

\bibliographystyle{apsrev4-1}
\bibliography{refs}

\appendix

\section{Exact coefficients for the open triangular lattice}
\label{sec:app-tri}

This appendix records the construction used to obtain the reference
coefficients of Sec.~\ref{sec:results-tri}.

\paragraph{Lattice.}
The reference graphs are open $L\times L$ patches of the triangular
lattice: vertices $(x,y)$ with $x,y\in\{0,\ldots,L-1\}$, and bonds
$(x,y)\leftrightarrow(x+1,y)$, $(x,y)\leftrightarrow(x,y+1)$, and
$(x,y)\leftrightarrow(x+1,y-1)$ whenever both endpoints lie in the
domain, with no periodic identification.  Hence $N = L^2$ and
\begin{equation}
\label{eq:tri-M}
M = 2L(L-1)+(L-1)^2 = 3L^2-4L+1 ,
\end{equation}
that is $M = 33,\,161,\,385,\,705$ for $L = 4,8,12,16$, matching
Table~\ref{tab:systems}.  This is the same finite graph on which the
sampled null-space bases are built.

\paragraph{Directed-walk determinant.}
Introduce the six directed steps
\begin{equation}
\begin{split}
&d_0=(1,0),\quad d_1=(0,1),\quad d_2=(-1,1),\\
&d_3=(-1,0),\quad d_4=(0,-1),\quad d_5=(1,-1),
\end{split}
\end{equation}
which, in the equilateral embedding $r(x,y)=x\,e_0+y\,e_{\pi/3}$ with
$e_\theta$ the unit vector at angle~$\theta$, point along the
directions $\pi\alpha/3$.  A directed state is a pair $(j,\alpha)$ of a
site and an attempted outgoing direction, so the turning matrix $M_1$
has dimension $6L^2\times6L^2$, with entries
\begin{equation}
M_1(j,\alpha;k,\omega)=
\begin{cases}
\phi(\zeta), & k=j+d_\alpha \text{ in the domain},\\
             & \zeta=(\omega-\alpha)\bmod 6,\\[2pt]
0, & \text{otherwise},
\end{cases}
\label{eq:M1}
\end{equation}
where the walk turns through the angle $\theta=\pi\zeta/3$ and
$\phi(\zeta)=\e^{\ii\theta/2}$ taken on the branch
\begin{equation}
\phi(\zeta)=
\begin{cases}
\e^{\ii\pi\zeta/6}, & \zeta=0,1,2,\\
0, & \zeta=3,\\
\e^{-\ii\pi(6-\zeta)/6}, & \zeta=4,5 ,
\end{cases}
\label{eq:phi}
\end{equation}
the zero at $\zeta=3$ forbidding immediate backtracking.  States whose
step leaves the domain are inactive and contribute trivial factors.
For this construction the Kac--Ward formula for finite planar
graphs~\cite{KacWard1952,Cimasoni2010} gives
Eq.~\eqref{eq:det-square-main},
$\det(I-vM_1)=G(v)^2$.  Appendix~B of Stanley~\cite{Stanley1971}
presents the corresponding directed-walk construction for the square
lattice; here the same construction is applied to the six directed edge
orientations of the triangular lattice, with the appropriate
turning-angle phases.

\paragraph{Evaluation.}
Let $s=6L^2$ and write the characteristic polynomial of $M_1$ as
$\det(\lambda I-M_1)=\lambda^s+c_1\lambda^{s-1}+\cdots+c_s$.  Its
coefficients are computed exactly by the Faddeev--LeVerrier
recursion~\cite{Gantmacher1959}: set $C_1=M_1$,
$c_1=-\operatorname{tr}C_1$ and, for $k=2,\ldots,s$,
$C_k=M_1(C_{k-1}+c_{k-1}I)$, $c_k=-\operatorname{tr}(C_k)/k$.  The same
coefficients give $\det(I-vM_1)=1+c_1v+\cdots+c_sv^s$.  Every nonzero
entry of $M_1$ is a twelfth root of unity, and in our implementation
every intermediate matrix element remains a single phase monomial
$a\,\zeta^{\,p}$ with $a\in\mathbb{Z}_{\ge0}$, $\zeta=\e^{\ii\pi/6}$
and $p\in\{0,\ldots,11\}$, a property checked at run time.  Exact
arithmetic therefore reduces to arbitrary-precision integer magnitudes
together with a phase index modulo~12.

Finally $\det(I-vM_1)$ is factorized exactly.  Every irreducible factor
occurs with even multiplicity, as guaranteed by
Eq.~\eqref{eq:det-square-main}, so the polynomial square root $G(v)$
is obtained by halving these multiplicities and fixing $G(0)=1$.  Its
coefficients are the nonnegative integers $k_l$.  As stated in
Sec.~\ref{sec:results-tri}, they satisfy $k_3 = 2(L-1)^2$ and
$\sum_l k_l = 2^{M-N+1}$; the latter follows because
$\sum_l k_l = |\ker_{\F_2}B| = 2^{M-\mathrm{rank}_{\F_2}B}$ and the
vertex--edge incidence matrix of a connected graph has
$\mathrm{rank}_{\F_2} = N-1$.

\section{The three-dimensional Baxter--Wu system}
\label{sec:app-bw3d}

This appendix records the counting behind Eq.~\eqref{eq:bw3d-M} and the
null-space dimensions of Table~\ref{tab:systems} for the geometry
defined in Sec.~\ref{sec:results-bw}.

\paragraph{Counting the faces.}
In cubic coordinates the triangular faces of $\mathcal G_L$ fall into
three Cartesian plane families ($xy$, $xz$, $yz$), each consisting of
$L$ planes carrying $2(L-1)^2$ triangles apiece, together with one
oblique family of $2(L-1)^3$ triangles; this decomposition is
exhaustive and disjoint, and we have verified it face by face for
$L \le 6$.  Hence
$M = 6L(L-1)^2+2(L-1)^3 = 2(L-1)^2(4L-1)$, giving
$M = 88,\,270,\,608,\,1150$ for $L = 3,4,5,6$.  The auxiliary
nearest-neighbor graph has $E_{\mathrm{nn}}=3L(L-1)(2L-1)$ two-body
edges, that is $90,\,252,\,540,\,990$; these define the faces and are
not interaction terms.

\paragraph{Null-space dimensions.}
Let $B\in\F_2^{N\times M}$ be the site--triangle incidence matrix.  For
each simulated size $L=3,4,5,6$ direct row reduction over $\F_2$ gives
$\mathrm{rank}_{\F_2}B = N$, i.e.\ full row rank, so
\begin{equation}
n = M-N = 7L^3-18L^2+12L-2 ,
\end{equation}
which evaluates to $61,\,206,\,483,\,934$.  At $L=6$ this gives
$n = 1150-216 = 934$, matching the exported basis of $934$ vectors of
length~$1150$.

\section{Supplementary figures}
\label{sec:appendix}

This appendix collects the mean-energy comparisons for all five systems,
the weight distributions for the three- and four-dimensional lattices,
and the energy-space densities of states used as the baseline in
Sec.~\ref{sec:results}.

\begin{figure*}[t]
\centering
\includegraphics[width=0.78\textwidth]{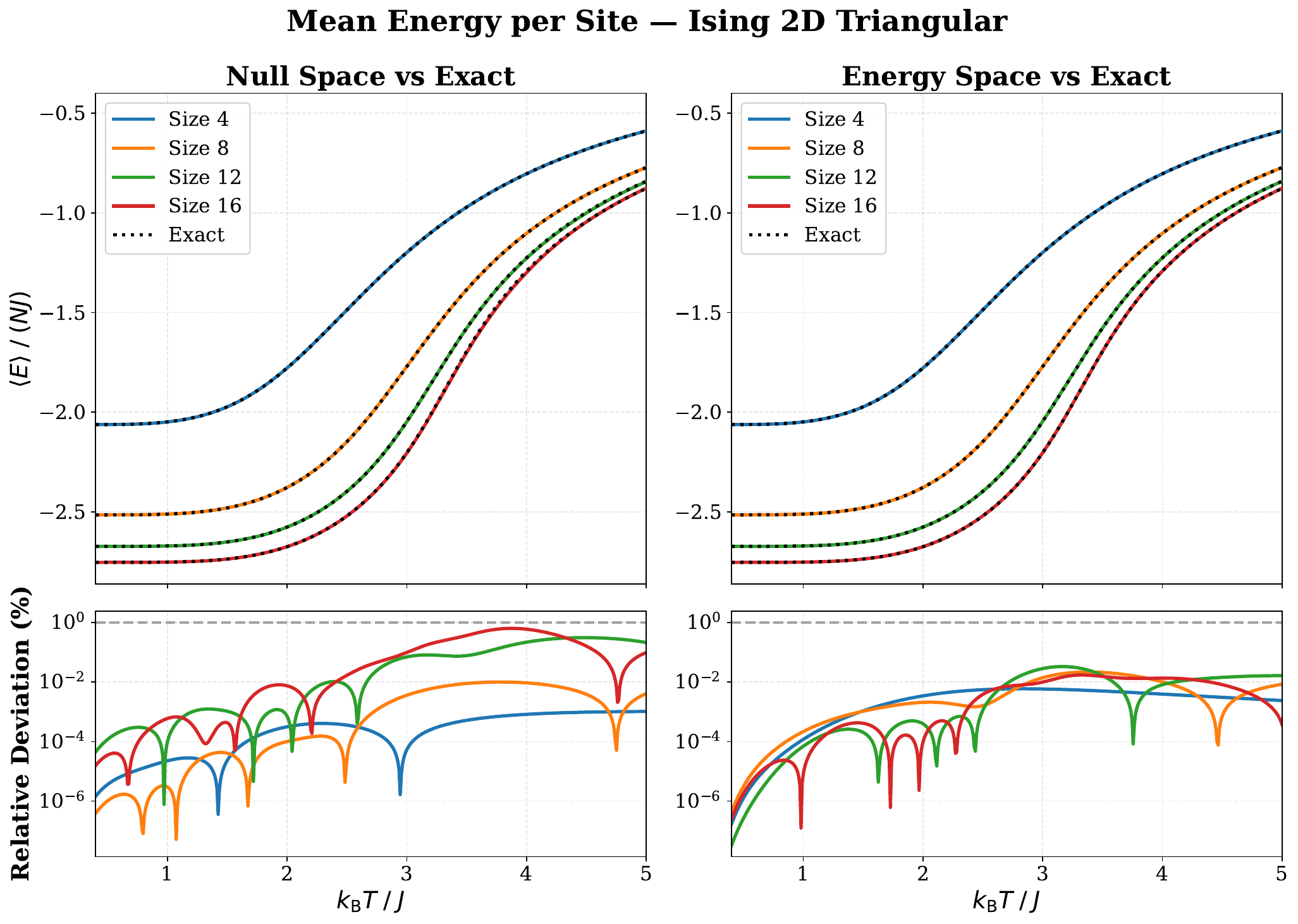}
\caption{Mean energy per site of the triangular-lattice Ising model from
cycle-space sampling (left) and energy-space sampling (right), each
against the exact finite-size result, with relative deviations below.}
\label{fig:app-energy-tri}
\end{figure*}

\begin{figure*}[t]
\centering
\includegraphics[width=0.48\textwidth]{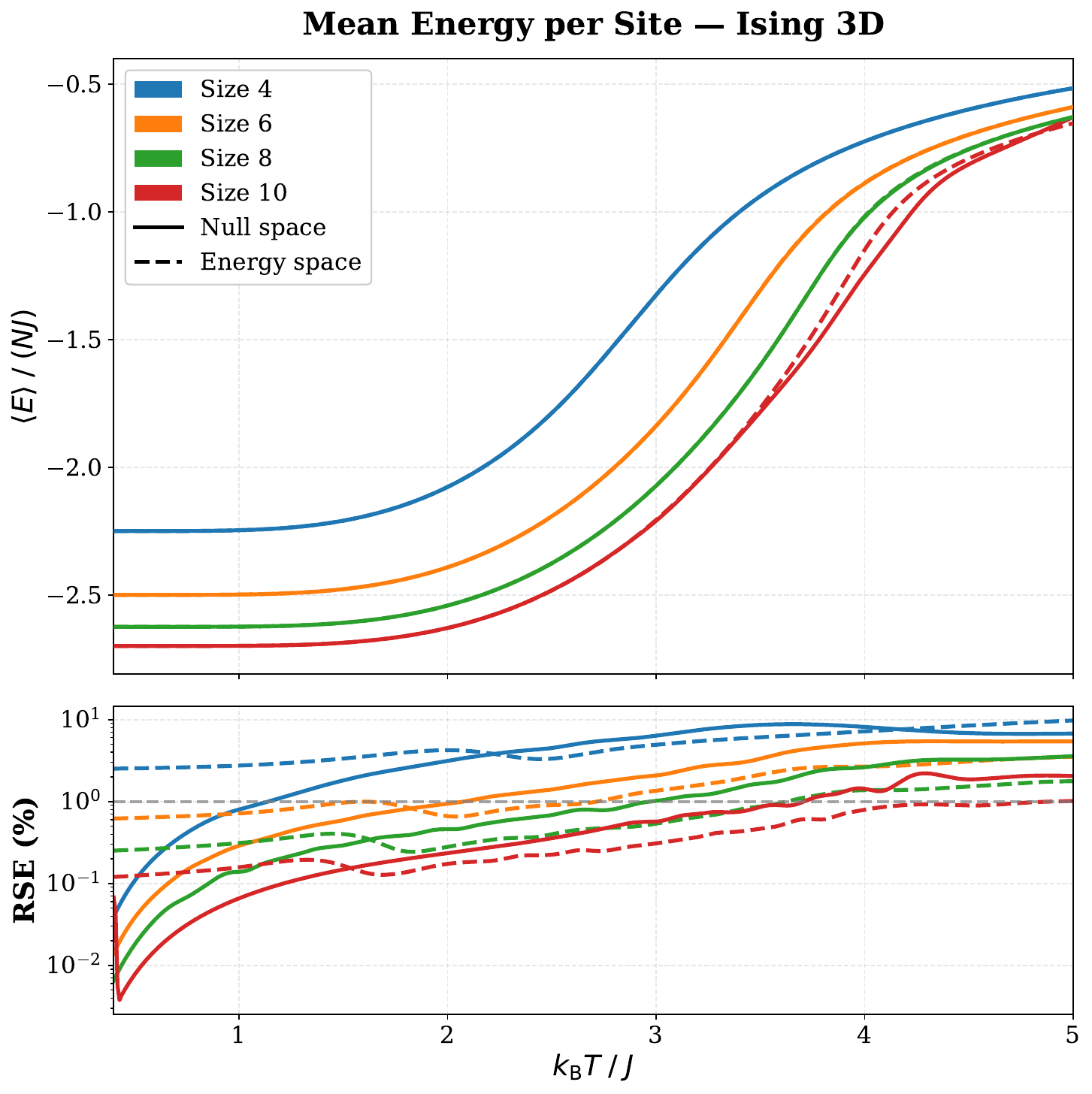}
\hfill
\includegraphics[width=0.48\textwidth]{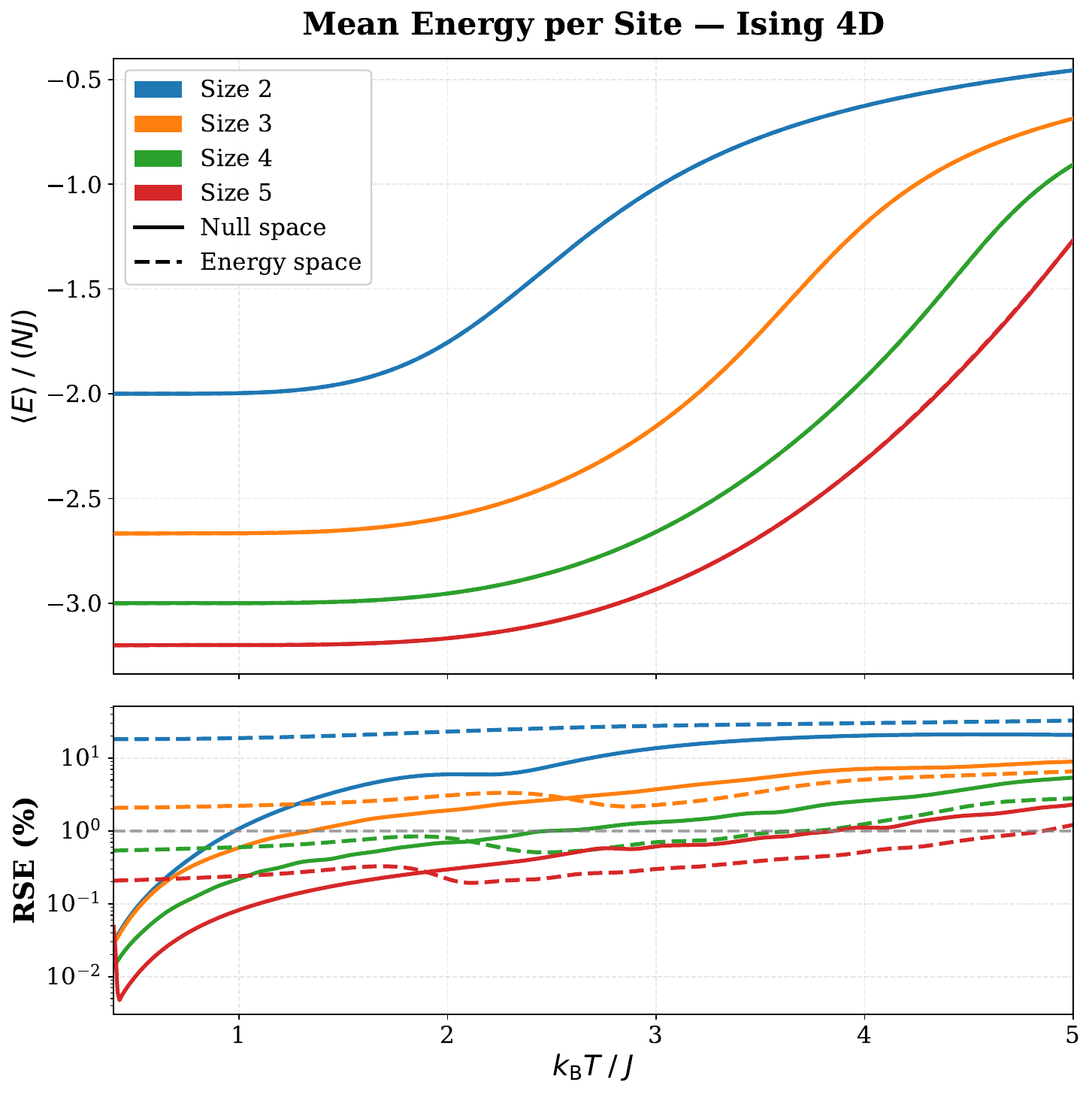}
\caption{Mean energy per site for the simple cubic (left) and
four-dimensional hypercubic (right) Ising models.  Solid curves are
cycle-space estimates, dashed curves energy-space estimates, with
relative standard errors below.}
\label{fig:app-energy-hd}
\end{figure*}

\begin{figure*}[t]
\centering
\includegraphics[width=0.48\textwidth]{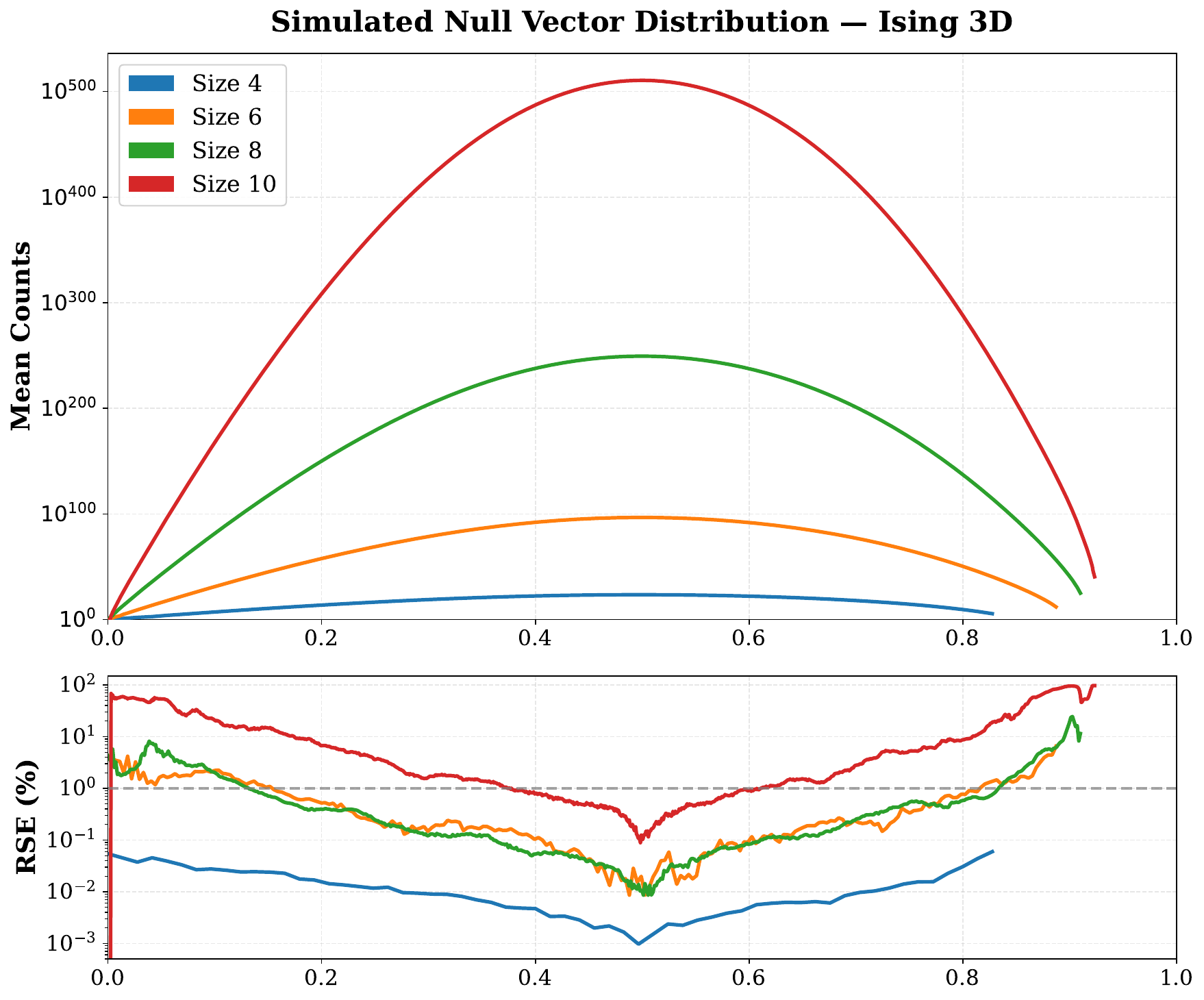}
\hfill
\includegraphics[width=0.48\textwidth]{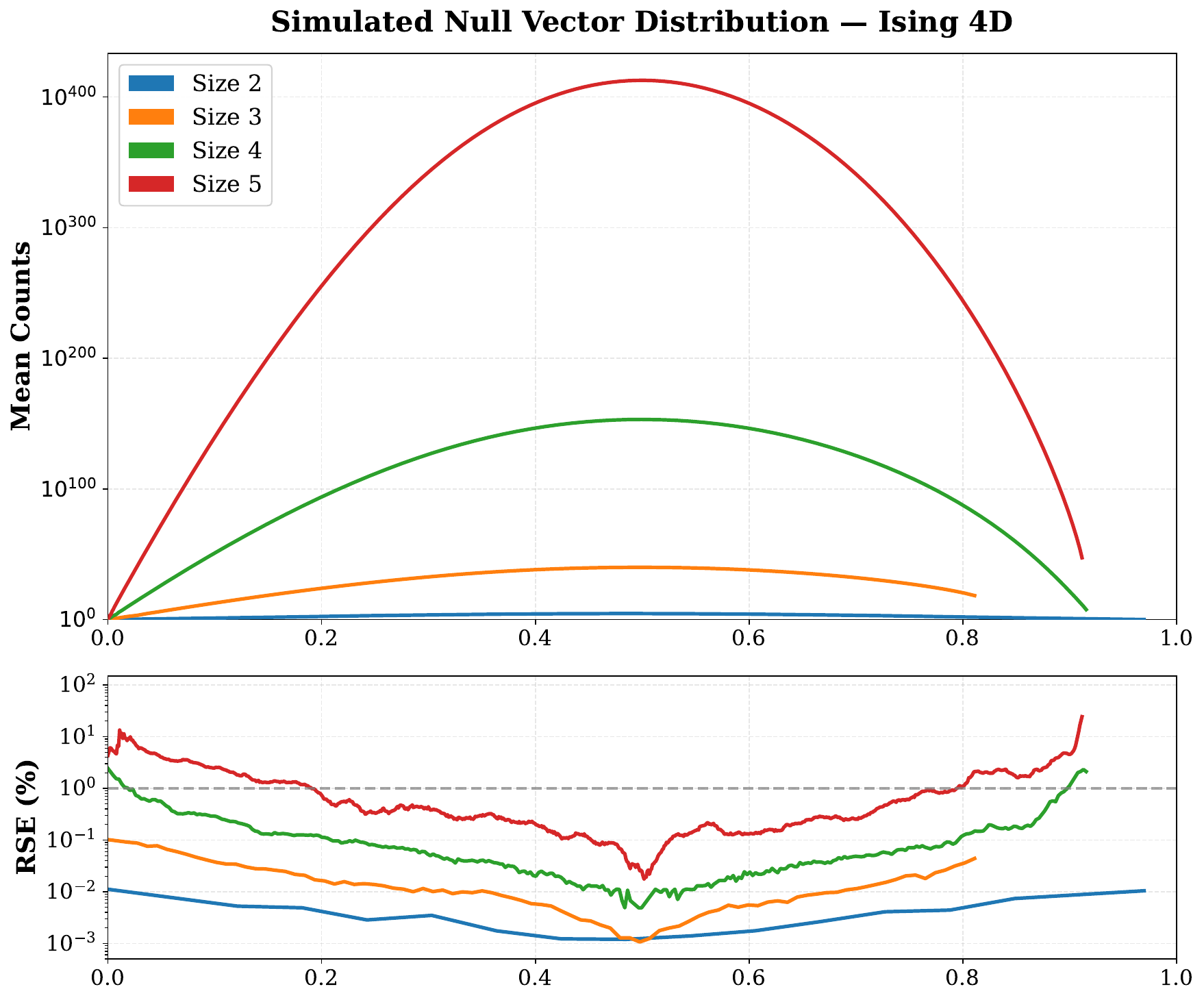}
\caption{Weight distributions of the cycle codes of the simple cubic
(left) and four-dimensional hypercubic (right) lattices, with relative
standard errors below, plotted against the normalized Hamming
weight $l/M$.  The distributions span roughly $500$ and $400$ decades
respectively at the largest sizes.}
\label{fig:app-null-hd}
\end{figure*}

\begin{figure*}[t]
\centering
\includegraphics[width=0.44\textwidth]{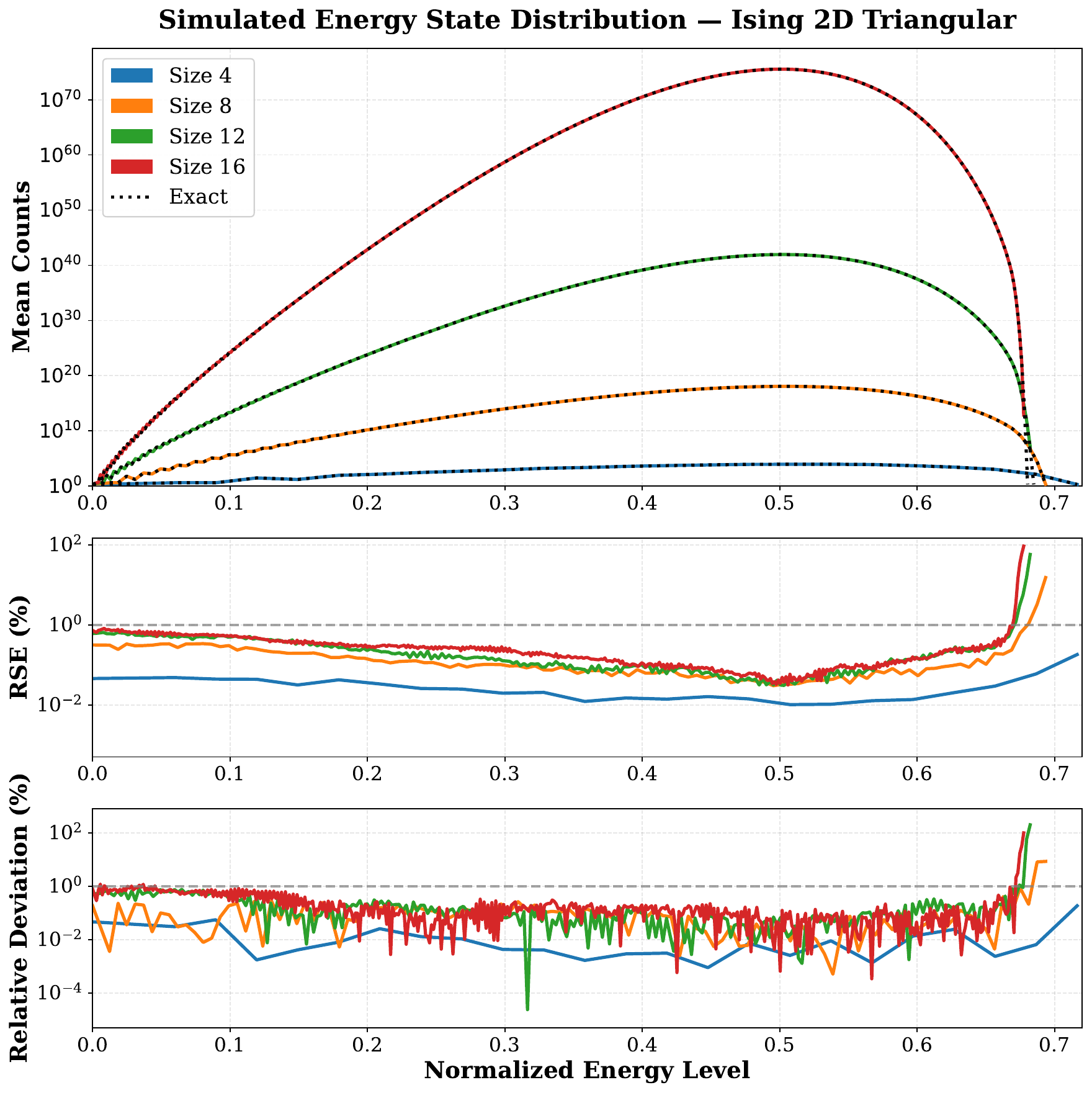}
\hfill
\includegraphics[width=0.44\textwidth]{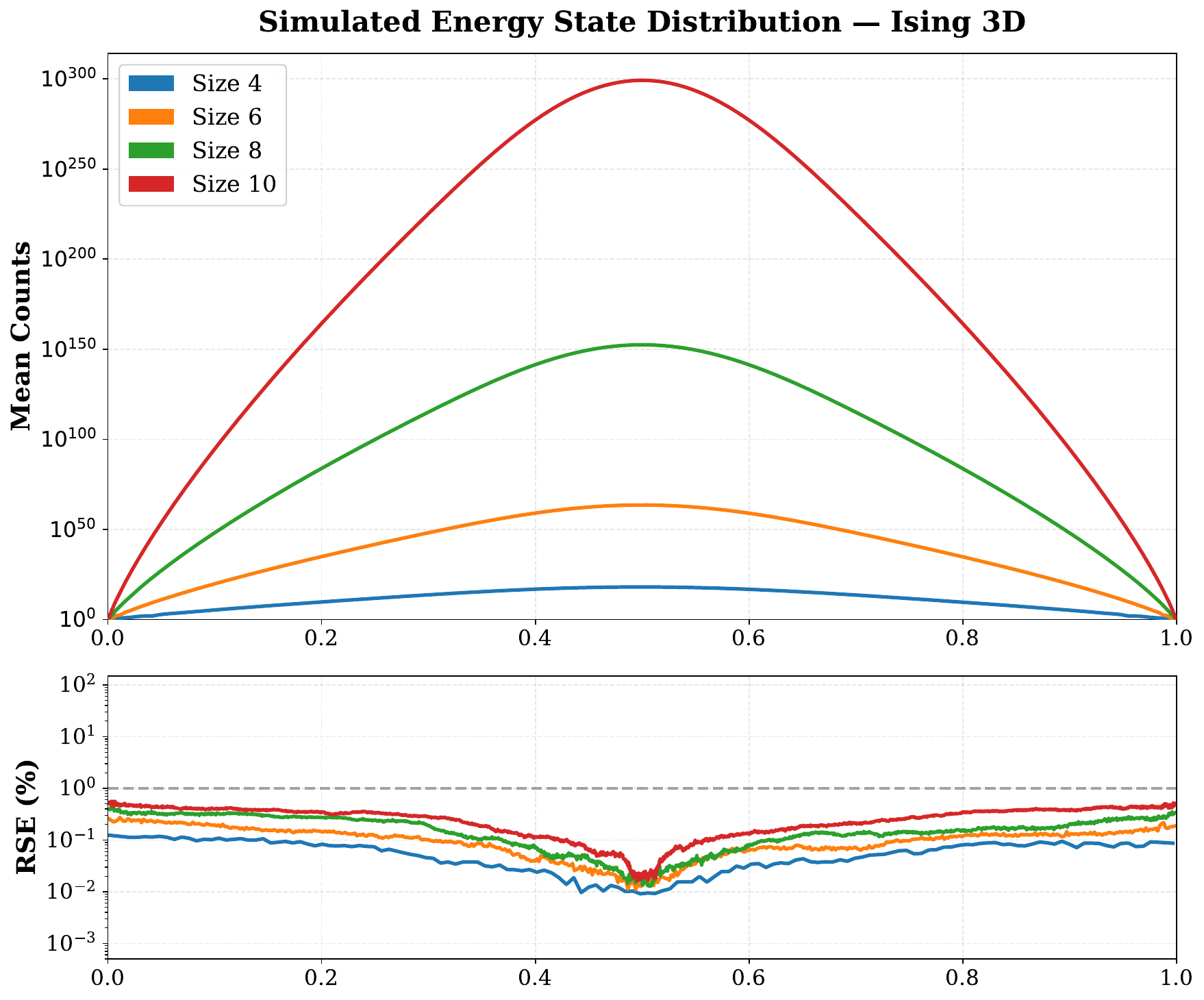}\\[1em]
\includegraphics[width=0.44\textwidth]{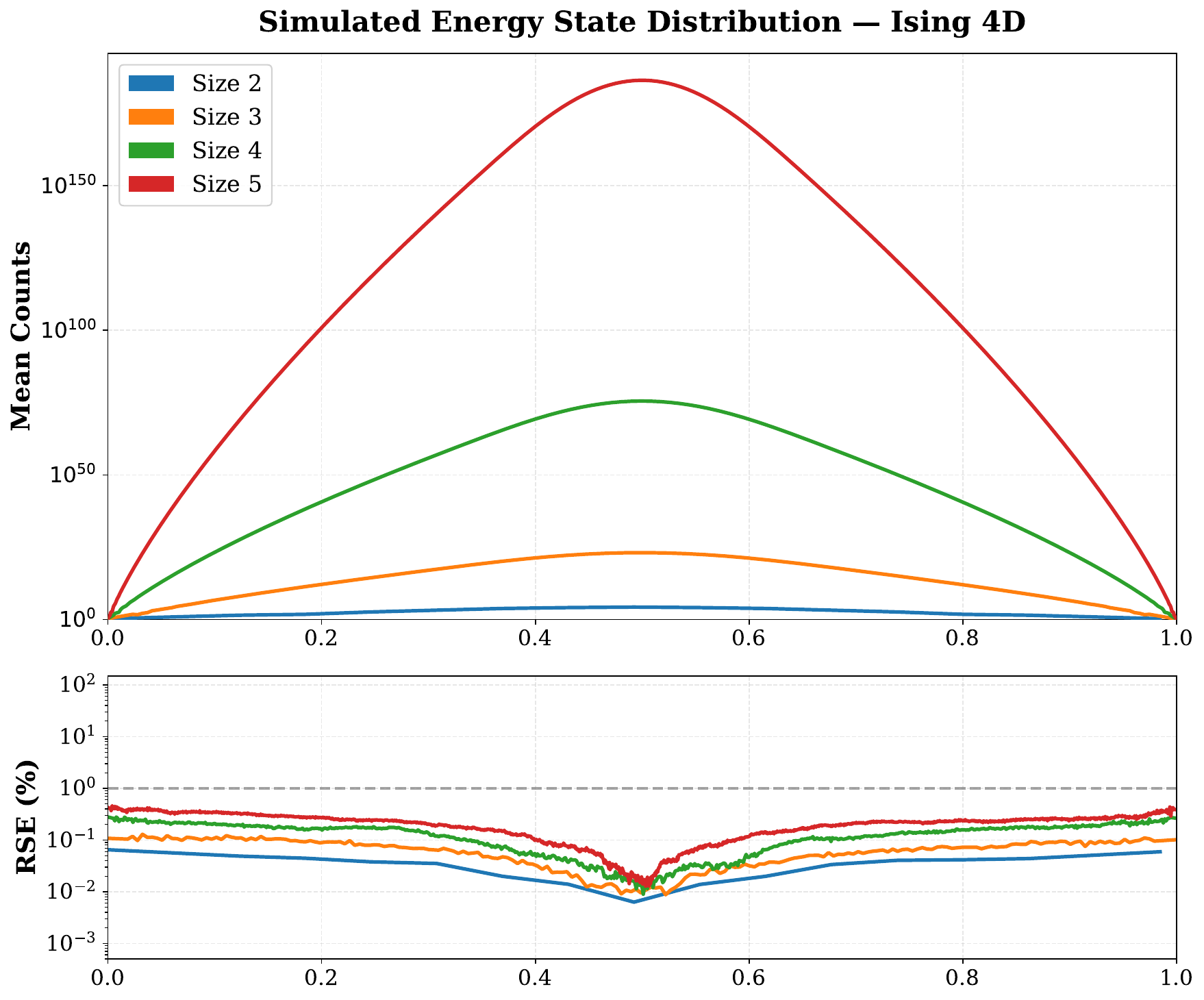}
\caption{Energy-space densities of states $g(E)$ obtained by
conventional Wang--Landau sampling, for the triangular (top left),
simple cubic (top right), and four-dimensional hypercubic (bottom)
lattices.  These are the baselines used in Figs.~\ref{fig:tri-cv},
\ref{fig:hd-cv}, \ref{fig:app-energy-tri}, and
\ref{fig:app-energy-hd}.}
\label{fig:app-edist}
\end{figure*}

\begin{figure*}[t]
\centering
\includegraphics[width=0.48\textwidth]{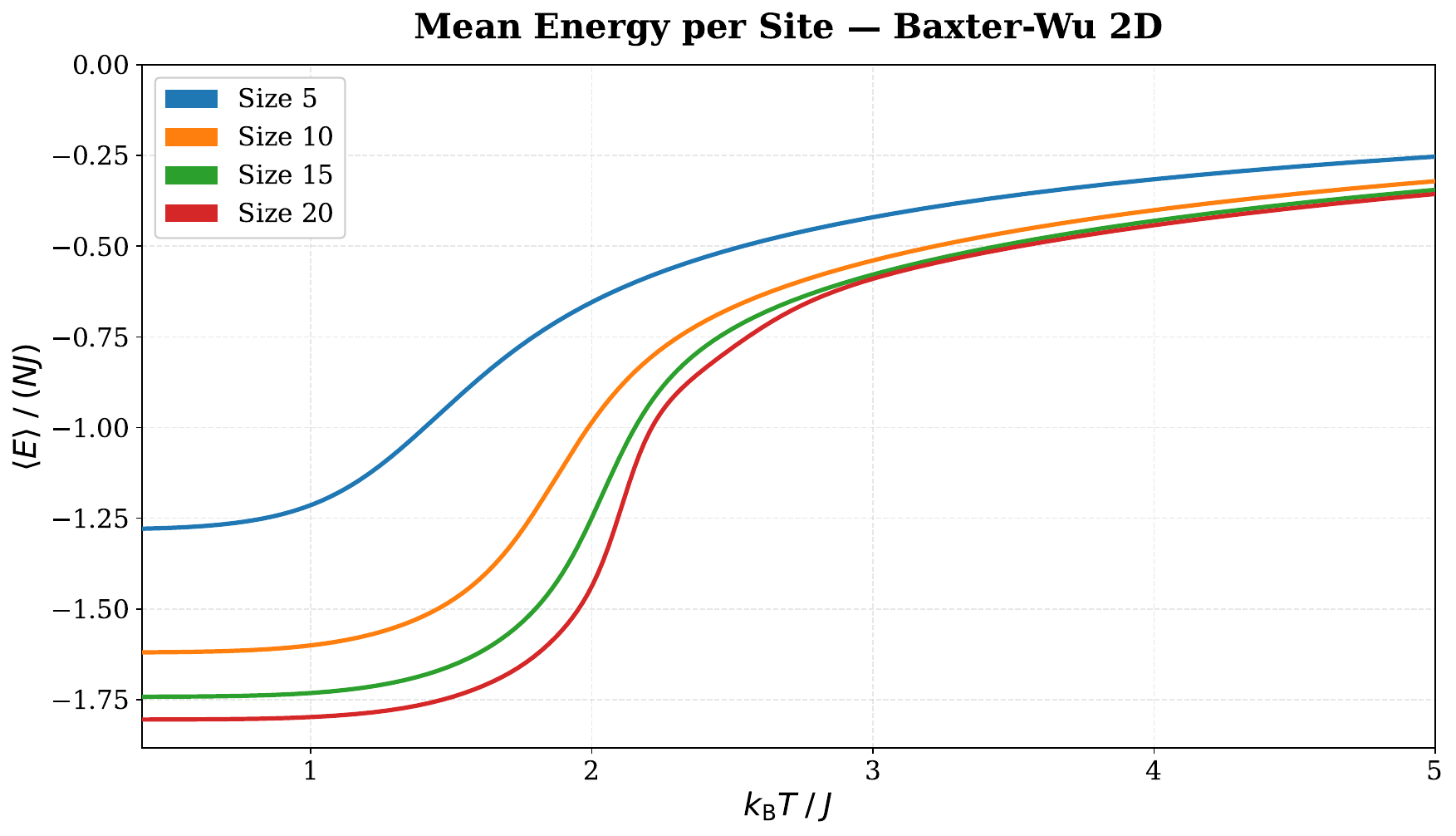}
\hfill
\includegraphics[width=0.48\textwidth]{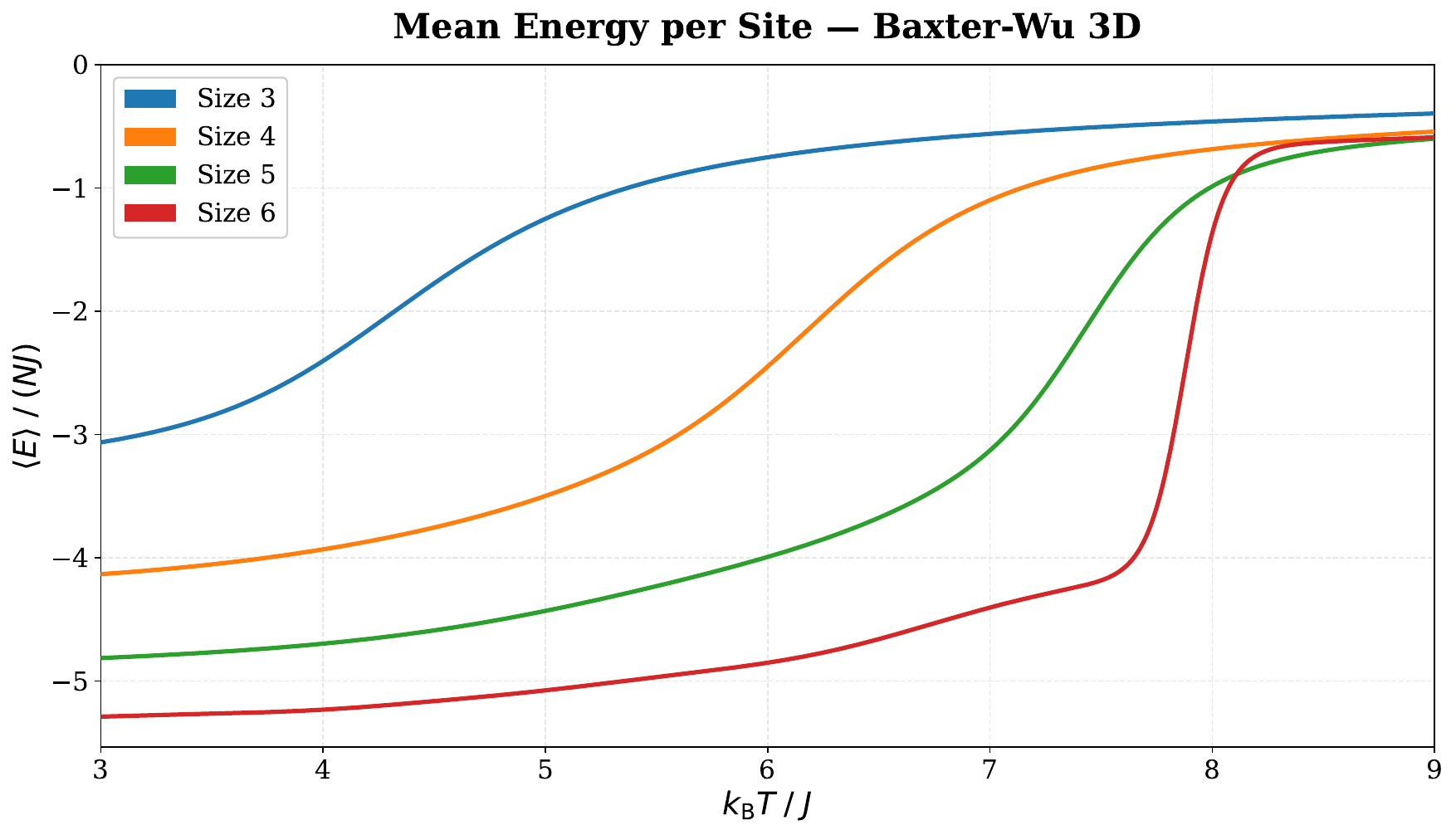}
\caption{Mean energy per site for the two-dimensional (left) and
three-dimensional (right) Baxter--Wu models from cycle-space sampling.
Both are monotonically increasing in temperature and approach the
ground-state value $-M/N$ at the low-temperature end, as required.  The
steep rise of the $L=6$ curve near $k_BT/J \simeq 7.9$ in the right
panel coincides with the specific-heat maximum of
Fig.~\ref{fig:bw-cv}.}
\label{fig:app-energy-bw}
\end{figure*}

\end{document}